\documentclass[12pt]{article}
\usepackage{amssymb}
\usepackage{amsfonts}
\usepackage[nohead]{geometry}
\usepackage[singlespacing]{setspace}
\usepackage[bottom]{footmisc}
\usepackage{indentfirst}
\usepackage{tikz}
\usepackage{pdflscape}
\usepackage{siunitx}
\usepackage{subcaption}

\usepackage{eurosym}
\usepackage[flushleft]{threeparttable}
\usepackage{graphicx}
\usepackage{rotating}
\usepackage{textcomp}
\usepackage{lscape} 
\usepackage{booktabs,caption}
\usepackage[flushleft]{threeparttable}
\usepackage{rotating}
\usepackage{xcolor}
\usepackage{upgreek}
\usepackage{tabularx}
\usepackage{lscape}
\usepackage{amssymb}
\usepackage{float}
\usepackage{amsfonts}
\usepackage{dsfont}
\usepackage[leqno]{amsmath}

\makeatletter
\def\@biblabel#1{\hspace*{-\labelsep}}
\makeatother
\usepackage{hyperref}
\hypersetup{
    colorlinks=true,
    linkcolor=blue,
    citecolor=blue,
    urlcolor=blue 
}

\usepackage{natbib}
\date{}

\begin{document}

\newgeometry{left=1.2in,right=1.2in,top=0.8in,bottom=0.8in} 

\title{University as a Melting Pot: \\ Long-term Effects of Internationalization} 
\author{Stanislav Avdeev\footnote{This version: March 2026. University of Amsterdam, Roetersstraat 11, 1018 WB Amsterdam, the Netherlands, e–mail: \href{mailto:stnavdeev@gmail.com}{stnavdeev@gmail.com}. I am grateful to Hessel Oosterbeek and Bas van der Klaauw for their extensive feedback. I also thank two anonymous referees, Adam Booij, Michel Beine, Leah Boustan, Thomas Buser, Monique de Haan, Sander de Vries, Bas Machielsen, Richard Murphy, Morgan Raux, Giuseppe Sorrenti, Dmitrii Shchetinin, Umut Özek, Alexander Yarkin, and participants at various workshops and conferences. I acknowledge financial support from a Doctoral Research Grant from the European Association for International Education. I thank Jeroen Kleingeld for his support with the data from the Nationale Studenten Enquête (NSS) provided by the Landelijk Centrum Studiekeuze. In this paper, I make use of data from the LISS panel (Longitudinal Internet Studies for the Social Sciences) managed by the non-profit research institute Centerdata (Tilburg University, the Netherlands). The non-public administrative micro data used in this paper are available via remote access to the Microdata services of Statistics Netherlands. The NSS and LISS survey data used in this paper must be requested from their respective data providers under their access conditions.} 
	\medskip \\ }
\maketitle
	
\thispagestyle{empty}

\sloppy 
	
\vspace{-1.5cm}
\begin{abstract}

	\noindent This paper provides the first evidence on the impact of exposure to international students on the long-term outcomes of native students. I combine unique survey and administrative data from the Netherlands covering one million students across three decades and employ an across-cohort design. I find that exposure to international students leads natives to \textit{(i)} form social ties with non-natives, \textit{(ii)} hold more positive attitudes towards migration and learning about other cultures, and \textit{(iii)} seek opportunities abroad. Notably, I find precisely estimated zero effects on employment, income, entrepreneurship, and the share of international co-workers up to 25 years after university entry.

	\strut
		
		
	\strut
		
		
\end{abstract}

\restoregeometry
	
\pagebreak
	
\onehalfspacing 
	
\section*{Introduction}
\label{introduction}

Over the past two decades, the number of students studying abroad has tripled globally \citep{Unesco2024}. Universities attract international students to increase revenue, expand social networks, and offer students opportunities to work with diverse peers in an international setting. However, politicians have raised concerns about the displacement of domestic students, negative labor market impacts, and the costs of subsidizing the education of foreign students. These concerns, along with broader debates on migration, have led to measures limiting student inflows in countries that host half of all international students, such as Australia, Canada, Denmark, Finland, France, Germany, the Netherlands, Norway, Sweden, Switzerland, the UK, and the US.\footnote{Since 2023, at least nine countries have implemented restrictive policies targeting international students. The Australian and Canadian governments have introduced annual caps on student visas \citep{Ross2024, Bhardwa2024}. Meanwhile, tuition fees for international students have increased in Finland \citep{Dixon2024b}, Germany \citep{Upton2023}, Norway \citep{Upton2023b}, and Switzerland \citep{Etias2024}. The Dutch government is currently reducing the number of English-taught bachelor’s programs \citep{Dixon2024}, and the UK has recently restricted international bachelor’s students from bringing family members \citep{Jack2023}. The US has cut funding for STEM research that supports international students and is considering a major travel ban \citep{Knox2025}. Prior to 2023, three other countries enacted similar measures. France \citep{Matthews2018} and Sweden \citep{Fearn2011} implemented increases in tuition fees for international students, while Denmark imposed caps on international student enrollments in 2021 but reversed the policy in 2023, acknowledging unintended negative consequences from restricting foreign student inflows \citep{Jack2023b}.} While these policies target international students, they may have unintended consequences for native students. 

This paper examines the impact of exposure to international students during the first year of a bachelor's program on the long-term outcomes of native students in the Netherlands. I employ an across-cohort design, following the approach introduced by \cite{Hoxby2000}, which is particularly well-suited to the Dutch context, where the absence of grade-based selection and the lack of fixed capacity constraints make the variation in the number of native and international students largely idiosyncratic. Specifically, I exploit cohort-to-cohort variation in the share of international students within university programs over time. My approach allows for student selection \textit{across} programs, as it does not compare students from different programs but rather compares students \textit{within} the same program \textit{across} different cohorts.

The identification strategy relies on the assumption that students do not select into specific cohorts \textit{within} a program based on the presence of international students. This assumption is plausible for two reasons. First, prospective students are generally unaware of the international student composition in their program until they begin their studies. Second, even if students could estimate the approximate composition, it is unlikely that they would alter their enrollment timing to avoid or seek out international peers. To support this assumption, I provide evidence that the share of international students is uncorrelated with the likelihood of native students taking a gap year, a rich set of pre-determined characteristics at the student, sibling, and parent levels, peer composition and indicators of program quality. 

The results show that exposure to international students increases natives' social ties with non-natives and their openness to migration. During university, a 10 percentage point increase in the share of international students leads to a 0.12 standard deviation increase in positive attitudes towards the internationalization of education and a 0.07 standard deviation increase in satisfaction with opportunities to learn about other cultures. Fifteen years after entering university, a 10 percentage point increase leads to \textit{i)} a 5.9\% increase in the probability of cohabiting with a non-native; \textit{ii)} a 4.2\% increase in the probability of marrying a non-native, and \textit{iii)} a 4\% increase in the probability of emigration. The results withstand a battery of robustness checks, including adjustments for multiple hypothesis testing, time trends, program size, individual and peer characteristics, different sample definitions, definitions of international students, and a placebo test.

To illustrate the size of the effects, consider a policy whereby a university increases the share of international students in each incoming bachelor’s program cohort by 10 percentage points, given an average program size of 137 students. On average, such an increase would result in 1 additional native student per program cohabiting with a non-native, 1 additional native emigrating for every 3 programs, and 1 additional native marrying a non-native for every 5 programs.

Having established positive effects of international exposure on social ties and openness to migration, I then examine whether the presence of international students affects native students’ labor market outcomes. Policymakers often voice concerns that international students might strain educational resources and reduce educational quality, potentially disadvantaging natives in the labor market. However, I find that exposure to international students shows precisely estimated zero effects on employment, income, entrepreneurship, and the share of foreign-born co-workers in the firms where natives are employed up to 25 years after university entry. These findings suggest that exposure to international students makes natives more culturally open and internationally oriented, without compromising their local economic success.

Social integration effects are strongest in collaborative and diverse environments and absent in more competitive ones. For example, STEM fields and programs with fewer female students, which tend to be more competitive, show no significant effects on social ties between native and international students. In contrast, non-STEM fields and programs with a more balanced sex composition, foster stronger social integration. Additionally, social ties between native and international students are observed only in larger programs since they offer broader networks and more opportunities for interaction. Exposure to EEA+ peers enhances social integration, likely due to shared cultural and institutional backgrounds. Interactions with non-EEA+ students, who face greater cultural and linguistic differences, do not significantly enhance social integration but increase satisfaction with opportunities to learn about other cultures.

Theoretically, two main mechanisms could explain the positive effects on social outcomes: meeting opportunities and preference shift. An important reason to distinguish between these mechanisms is that they imply different patterns of general equilibrium effects. Under the meeting opportunities mechanism, the impact of exposure operates only within the university context and does not fundamentally change cultural attitudes. The reach of the policy is therefore narrow and bounded: only students who directly interact form social ties, which fade once students leave the university environment. In contrast, the preference-shift mechanism can generate more durable effects: natives exposed to diversity at university may remain culturally open long after graduation. From a policy perspective, this distinction is critical: while the meeting opportunities mechanism might yield short-lived and localized improvements, a shift in preferences can lead to broad and lasting changes in societal openness and integration.

Empirically, I find strong evidence supporting the preference-shift mechanism. If the meeting opportunities mechanism were the sole explanation, intercultural relationships would predominantly form among peers within the same university program. However, natives frequently establish intercultural ties extending beyond their immediate academic environment, developing relationships with non-natives outside the university. Moreover, exposure to international students significantly alters natives' attitudes: it increases favorable views towards the social integration of foreigners and support for European unification, while reducing negative attitudes towards foreigners moving into the neighborhood. These empirical patterns strongly suggest that exposure to international peers induces meaningful shifts in underlying preferences towards migration and cultural diversity, rather than merely providing convenient meeting opportunities for intercultural contact.

This paper expands existing research by providing new evidence in two areas: \textit{(i)} the impact of internationalization of higher education and \textit{(ii)} the relevance of the contact hypothesis in the context of university education.

First, I contribute to the small but growing literature on the impact of the internationalization of higher education. While most existing studies focus on academic and short-term labor market outcomes, this paper is the first to examine the impact of international peers on native students’ social outcomes and long-term labor market outcomes.\footnote{Related studies have explored crowd-out effects \citep{Borjas2004, Shen2016, Machin2017, Shih2017, Chen2023} and the resource implications of internationalization \citep{Zhu2023}. A large body of work has used Hoxby-style designs to study peer effects on academic outcomes, for example in the US \citep{Whitmore2005, Bifulco2011} and Israel \citep{Lavy2011, Lavy2012}. More recently, researchers have applied this approach to outcomes beyond academic achievement. These include labor market outcomes in Norway \citep{Black2013}, the US \citep{Bifulco2014, Carrell2018, Merlino2019, Merlino2024}, and Denmark \citep{Brenoe2020}; inter-ethnic family formation and neighborhood and workplace diversity in the US \citep{Merlino2019, Merlino2024} and Sweden \citep{Holmlund2023}; and attitudes towards minority groups in the US \citep{Merlino2019, Merlino2024}.} For example, \cite{Anelli2023} and \cite{Rakesh2023} examine the effects of exposure to foreign classmates in introductory first-year math courses at two US universities. \cite{Anelli2023} find that exposure to foreign students decreases the likelihood of native students graduating with a STEM degree and has no significant impact on expected earnings. Similarly, \cite{Rakesh2023} finds that exposure to foreign students lowers the graduation rate of native students. In the UK, \cite{Chevalier2020} find no impact of exposure to non-English speakers on the academic performance or emigration of native students six months after graduation. Likewise, \cite{Costas2023} find no effects of exposure on graduation, employment, and earnings six months post-graduation in the UK. To summarize previous findings alongside my own results, I estimate a meta-analytic pooled estimate of 0.0\%, with a 95\% confidence interval of $[-0.002, 0.002]$. Taken together, this indicates no impact of exposure to international students on the educational and labor market outcomes of natives. However, this paper demonstrates that social and cultural dimensions are first-order effects of internationalization, underscoring the importance of moving beyond conventional academic and economic metrics when evaluating its impact.

Second, this paper contributes to the literature on the contact hypothesis, which suggests that contact with minority groups can reduce prejudice.\footnote{The hypothesis posits that cross-cultural interactions can effectively break down stereotypes and prejudices between groups when a set of conditions, such as common goals, intergroup cooperation, support by social and institutional authorities, and equal status, are satisfied \citep{Williams1947, Allport1954}.} As universities become more diverse, they offer a valuable setting for shaping natives' preferences through everyday intercultural interaction. Previous studies in higher education have explored the impact of being assigned a roommate of a different race on inter-racial prejudice and connections in the US \citep{Boisjoly2006, Marmaros2006, Camargo2010, Carrell2019} and South Africa \citep{Corno2022}.\footnote{Related work investigates interventions in primary and secondary schools, including \cite{Rao2019} in India, \cite{Merlino2019} in the US, and \cite{Alan2021, Boucher2021} in Turkey.} This study is the first to test the contact hypothesis in a European university context. Moreover, by leveraging variation in group composition rather than relying on roommate assignments, this paper examines a more scalable and policy-relevant form of exposure.

My findings have significant policy implications for current debates on internationalization. Governments in a dozen countries are implementing policies to limit the number of international students. Based on the results of this paper, reducing international student enrollment could weaken social integration, cultural openness, and international opportunities for native students. This is especially relevant, as it affects students during their impressionable years, a critical period when attitudes and preferences are most adaptable. 

The remainder of the paper is structured as follows. Section \ref{background} provides an overview of the Dutch higher education system. Section \ref{design} describes the administrative and survey data, along with the identification strategy. Section \ref{results} presents the main results, a heterogeneity analysis, robustness checks, and a meta-analysis. Section \ref{conclusion} concludes.

\section{Institutional background}
\label{background}

Under the Dutch law, students who complete the academic track in secondary education are eligible for admission to any of the 13 research universities without grade-based selection criteria.\footnote{There are a few exceptions in programs like medicine, dentistry, and veterinary (see \cite{Leuven2013} and \cite{Ketel2016} for more details).} This policy ensures that the influx of international students does not displace Dutch students or affect their access to university education. Under the European Union (EU) law, all EU nationals must meet the same criteria.\footnote{These rules also apply to nationals of Iceland, Liechtenstein, Norway, and Switzerland.} This creates an ideal setting for my identification strategy, as year-to-year variation in native and international student numbers is unpredictable for both incoming students and university programs.

\begin{figure}[!htb]
	\caption{The share of international students enrolled in bachelor's programs at Dutch research universities in 1988--2019}
	\includegraphics[width=\textwidth]{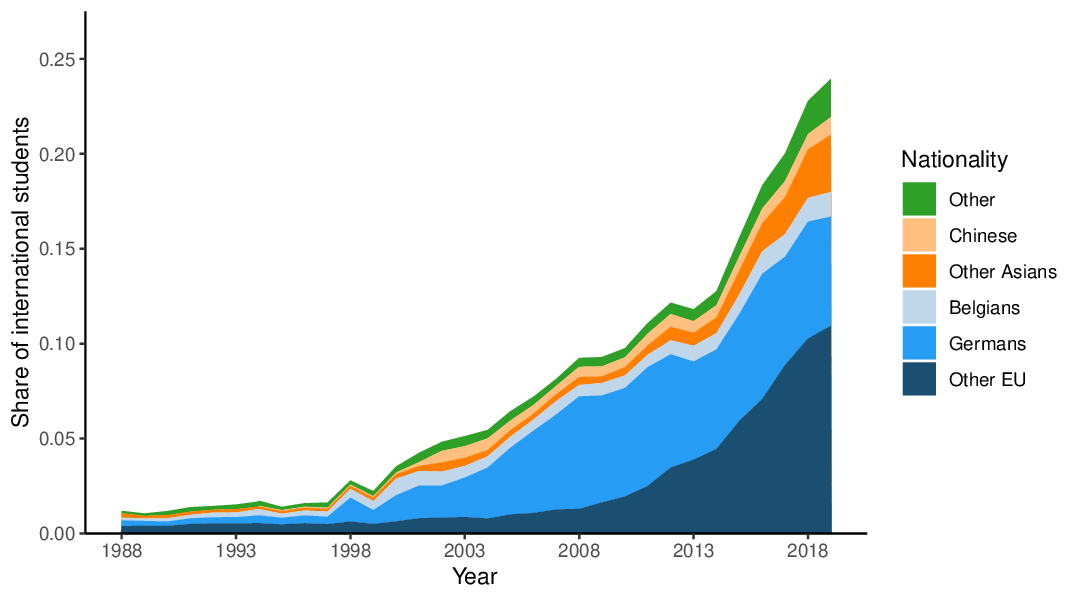}
	\label{trend}
	{\footnotesize \textit{Notes:} EEA+ refers to students from the European Economic Area (the EU, Iceland, Liechtenstein, and Norway) plus Switzerland. \\ 
    \textit{Source:} Calculated by the author using data from Statistics Netherlands.}
\end{figure}

There are low barriers for international students to enroll, because many programs use English as the primary language of instruction. Moreover, the tuition fees in the Netherlands are not high compared to those in other popular English-speaking destinations, such as Australia, the UK, and the US. Faculty salaries are determined by collective labor union agreements, ensuring a consistent level of quality across institutions. This commitment to accessibility and quality has resulted in all universities consistently ranking among the top institutions globally \citep{THE2023}. As a result of these factors, the share of international students in the Netherlands has seen a remarkable increase, rising from a modest 3\% in the 1990s to over 20\% by 2019 (see Figure \ref{trend}).

At Dutch universities, students enroll in specific programs from the first year. The first year consists of compulsory courses, ensuring that students in the same program share a common curriculum and interact regularly in peer groups. In later years, students may select elective courses, allowing for more specialization in their field of study. This paper focuses on the first year, where shared compulsory courses provide structured opportunities for peer interaction between native and international students.

Figure \ref{trend} shows that around 75\% of all international first-year bachelor's students originated from other European countries, with German and Belgian students forming the largest groups. Asian students constituted 16\% of the first-year international student body, with the largest groups being Chinese and Indian.

The share of international students differs substantially across fields of study. In 2019, Economics, Social Sciences, and Humanities had the highest shares of international students, around one-third. STEM fields followed with around 20\%, while Law and Health had the lowest percentages at around 10\%. The lower percentages in Law and Health are primarily due to language barriers, as most programs in these fields are taught in Dutch.

Universities have several incentives to attract international students. In the Netherlands, government funding is tied to Dutch and EEA+ student enrollment, making international recruitment financially advantageous. Moreover, non-EEA+ students pay higher tuition fees than their Dutch or EEA+ counterparts, providing additional financial boost. In 2019, tuition fees for Dutch and EEA+ national students were around \euro 2,000 per year, while fees for non-EEA+ nationals were approximately \euro 12,000. Additionally, university rankings frequently consider the share of international students and faculty, enhancing the appeal of a diverse student body.

\section{Research design}
\label{design}

In this section, I discuss the data sources and my identification strategy.

\subsection{Data}

\paragraph{Administrative data.} I utilize nationwide administrative data provided by Statistics Netherlands (CBS), which cover higher education enrollment from 1988 to 2023, secondary education enrollment from 2006 to 2023, tax registry from 2003 to 2023, firm registry from 2006 to 2023, welfare registry from 2003 to 2023, cohabitation registry from 1994 to 2023, and marriage registry from 1994 to 2023 \cite{CBS_GBAPERSOONTAB, CBS_GBANATIONALITEITBUS, CBS_GBAHUISHOUDENSBUS, CBS_PARTNERBUS, CBS_GBAVERBINTENISPARTNERBUS, CBS_HOOGSTEOPLTAB, CBS_KINDOUDERTAB, CBS_GBAOVERLIJDENTAB, CBS_INPATAB, CBS_IPI, CBS_POLISBUS, CBS_SPOLISBUS}. All registries are linked using unique identifiers. The education registry includes details on secondary school, exam grades, university, program, study duration, dropout, and graduation. The tax registry provides information on employment, entrepreneurship, and earnings. The firm registry contains all firms and their workers. The welfare registry provides information on unemployment and social benefits. The cohabitation and marriage registries include data on flatmate and spouse pairs. Additionally, I have access to demographic data such as sex, age, nationality, country of birth, neighborhood, residency, and death. I also link information on parents and siblings for balancing checks.

\paragraph{Survey data.} I merge two survey datasets: National Student Survey (NSS) and Longitudinal Internet Studies for the Social Sciences (LISS). The NSS panel includes questions on attitudes towards internationalization \citep{NSS2025}. The survey has been running since 2010 and contains aggregated responses from first-year students by program, university, and year of enrollment. These survey data are linked to administrative data using identifiers based on program, university, and year of enrollment.

The LISS panel includes questions on attitudes towards migration. The LISS panel is a nationally representative annual household survey that has been running since 2007 \citep{LISS2025}. The survey contains individual responses to a wide range of topics related to economic, social, and health outcomes. It consists of around 5,000 households, comprising approximately 7,500 individuals, drawn from a random sample of Dutch households. Since the LISS panel is representative of the full Dutch population, only a subset of survey respondents can be matched to the sample of university students. Approximately 15\% of the Dutch population holds a university degree, which corresponds to the share of LISS respondents that could be merged with the administrative data of university students. The merging is done using the same unique identifier that links all registries. The average age of respondents is 30.7.

\paragraph{Sample construction.} I focus on full-time students enrolled in bachelor's programs between 1988 and 2019. From the international student sample, I exclude exchange students (0.5\% of the sample) due to their short period of study. From the native student sample, I exclude first-generation immigrants (5.5\%) due to their uncertain status (whether they should be considered as native or international) and students older than 30 at the time of enrollment (1.5\%) as they do not represent the usual undergraduate populations. The results remain robust to the inclusion of first-generation migrants and students older than 30, as reported in the Appendix.

I define a native student as someone who holds Dutch nationality. To define an international student, I apply two criteria: not having Dutch nationality at the time of enrollment and not having completed secondary education in the Netherlands. This definition aims to capture individuals who lack prior exposure to the Dutch education system. I discuss the consequences of using alternative definitions of international students in the Robustness section. 

The analysis focuses on the first enrollment of students in bachelor's programs. This approach ensures accurate assignment of the treatment variable, defined as the share of international students at the time of a student's first enrollment. For instance, if a student enrolls in a program, subsequently drops out, and later enrolls in a different program, they are included in the study only once.

One limitation of the data is the lack of information on tutorial group assignments within programs. While students attend lectures together, tutorials are held in smaller groups. This allocation is handled centrally and typically does not allow for student choice. Depending on administrative logistics, students may remain with the same group across all courses or be reassigned for each course or block. Importantly, there is generally no input from students, parents, or faculty. As such, there is little scope for selection into classrooms within a program. 

\subsection{Identification strategy}

The aim of this paper is to estimate the impact of exposure to international students on the long-term outcomes of native students. Building on the model specifications of \cite{Carrell2018} and \cite{Anelli2023}, I exploit idiosyncratic variation in the share of international students within a bachelor's program over time, using the following model:
\begin{align}
	\label{model_main}
	Y_{ifut} = \beta D_{fut} + \alpha_{fu} + \tau_{ft} + \varepsilon_{ifut}
\end{align}
where $Y_{ifut}$ measures an outcome of interest\footnote{Appendix A provides a detailed description of the variable construction.} of a native student $i$ in a field of study $f$ at a university $u$ enrolled in a year $t$; $D_{fut} = \frac{Int_{fut}}{N_{fut} - 1}$ is the share of international students; $\alpha_{fu}$ are the field of study $\times$ university fixed effects; $\tau_{ft}$ are the field of study $\times$ year of enrollment fixed effects. Field of study refers to disciplines such as Economics, Business Administration, Sociology, etc. In total, there are 199 fields of study, 13 universities, and 32 years. For brevity, I refer to $\alpha_{fu}$ as program fixed effects and to $\alpha_{ft}$ as cohort fixed effects. I cluster standard errors at the intersection of field of study and university level as the treatment is assigned at this level \citep{Abadie2023}. Additionally, I report $q$-values using the method proposed by \cite{Anderson2008} to account for multiple hypothesis testing.

The inclusion of program fixed effects accounts for time-invariant characteristics, such as the location of the program, which could influence student selection across programs. Cohort fixed effects control for time-varying factors that affect all students, such as changes in tuition fees. I check the robustness of the main findings against alternative model specifications that include linear and quadratic field specific time trends, program size, and individual and peer characteristics. The results are robust to the inclusion of these controls and are reported in the Appendix.

\subsection{Identifying variation}

By leveraging the institutional setting of Dutch and EU law and incorporating program and cohort fixed effects, the primary source of variation in Dutch students' exposure to international peers arises from differences in cohort entry, which are largely driven by birth years. This variation arises from fluctuations in both the number of native students and the number of international students within programs.

First, the number of native students varies due to natural fluctuations in Dutch birth cohorts, leading to random changes in university enrollment sizes. Second, the number of international students fluctuates for two reasons. Admission offices at Dutch universities have no control over the number of both Dutch and EEA+ students, resulting in some randomness influenced by population changes in other European countries. Additionally, various push factors from EEA+ and non-EEA+ countries further contribute to these fluctuations. In summary, the variation in the number of both native and international students entering Dutch universities is largely driven by natural cohort fluctuations, external demographic changes, and push factors from other countries, creating a degree of randomness in student inflow.

As discussed in \cite{Angrist2014}, there needs to be sufficient variation in the cohort-to-cohort changes of the share of international students. Figures \ref{share} and \ref{share_residualised} in the Appendix illustrate the variation in the share of international students, both before and after accounting for the fixed effects.\footnote{There remains meaningful residual variation in the samples merged with the NSS and LISS panels: the standard deviation of the residualized share is 3.9 percentage points in the NSS sample and 3.2 percentage points in the LISS sample (see Figures \ref{share_nss} and \ref{share_residualised_nss}  for the NSS sample). Due to privacy regulations, histograms for the LISS sample cannot be plotted, as the cell sizes are small.} Even after controlling for program and cohort fixed effects, there is still enough variation to precisely estimate the parameter of interest. Specifically, one standard deviation of the share of international students is 12.6 percentage points, while the residualized share has a standard deviation of 5.9 percentage points. To put this in perspective, an increase of one standard deviation in the residualized share translates to adding approximately 8 international students to an average program of 137 students. This year-to-year variation is largely due to Dutch universities having no control over the number of incoming Dutch and EEA+ nationals due to the Dutch and EU laws, leading to idiosyncratic fluctuations in the share of international students.

\subsection{Identification challenges}

In this section, I discuss three challenges to my identification strategy: selection, reflection, and common shocks.

\paragraph*{Selection.} My identification strategy does not rule out student selection \textit{across} programs. For example, if students know from previous years that there are many international students in Economics at Maastricht University due to its proximity to the border with Germany and Belgium, and fewer international students in Economics at Erasmus University, and they decide where to enroll based on this knowledge, this is not a violation of my identification strategy. This is because I do not compare students across different programs (Economics at Maastricht University vs. Economics at Erasmus University). Instead, I compare students \textit{within} the same program \textit{across} different cohorts (Economics at Maastricht University in 2010 vs. Economics at Maastricht University in 2011).

The identification strategy relies on the assumption that students do not select into cohorts \textit{within} a program based on the presence of international students. For example, if students decide to enroll in Economics at Maastricht University in 2011 and not in 2010 because of the different share of international students, then this violates my identification strategy. This is unlikely due to two reasons. First, prospective applicants do not observe the composition of students in the year they enroll until their first day in the classroom. Only when students enroll in a program and start their first day can they observe this composition. That is why I measure the share of international students and all other peer outcomes at the start of the academic year. For example, if some students enroll after the start of the academic year, I do not include them in my sample, and they are not included to measure peer characteristics.

Second, even if students could anticipate the approximate composition, they would have to react to that information to violate my identification strategy. In other words, if students delay (enroll next year instead of this year) or accelerate (enroll this year instead of next year) their enrollment into a specific program based on the share of international students in this program, then it would violate this assumption. There are two cases of how students can do that. Imagine that students can observe the approximate composition of their peers and decide to either delay or accelerate their enrollment. To delay enrollment, students who would not have otherwise taken a gap year choose to take one. Conversely, to accelerate enrollment, students who would have otherwise taken a gap year decide to skip it.

To examine whether native students react to the share of international students, I look at whether the probability of taking a gap year changes, as well as a battery of other observable characteristics. I test for selection using balancing tests, regressing observable characteristics on native students’ exposure to international students, using the following model:
\begin{align}
	\label{model_balance}
	X_{ifut} = \beta D_{fut} + \alpha_{fu} + \tau_{ft} + \varepsilon_{ifut}
\end{align}
where $X_{ifut}$ includes student, sibling, and parent characteristics.

Table \ref{balance_1} reports estimates with and without program and cohort fixed effects. Without fixed effects, many estimates are statistically significant. With fixed effects, only 1 of 32 estimates remains significant at the 1\% level. I do not reject the null hypothesis that the coefficients are jointly equal to zero. Consistent with the scenario discussed above, students are neither more nor less likely to take gap years in response to the presence of international students. It is worth noting that the lack of statistical significance is not due to large standard errors but rather due to the small size of both the coefficients and the standard errors.

To assess the possibility of serial selection across cohorts, where students base their enrollment decisions on the international student composition of previous cohorts, I test whether the share of international students in year $t-1$ predicts the observable characteristics of Dutch students in year $t$. Table \ref{balance_2} shows that only 4 of the 32 estimates is statistically significant at the 10\% level. I do not reject the null hypothesis that the coefficients are jointly equal to zero.

Further, I regress the characteristics of peers of native students in a program on the share of international students in that program, using the following model:
\begin{align}
	\label{model_common_shocks}
	\bar{X}_{fut} = \beta D_{fut} + \alpha_{fu} + \tau_{ft} + \varepsilon_{fut}
\end{align}
where $\bar{X}_{fut}$ includes student, sibling, and parent characteristics collapsed to the program level. Table \ref{balance_3} shows that only 1 of 32 estimates is significant at the 10\% level. Again, I do not reject the null hypothesis that the coefficients are jointly equal to zero.

Table \ref{balance_4} examines the correlation between the share of international students and indicators of program quality. As proxies for program quality, I use the program size and the outcomes of graduates from the same year the share of international students is measured. I regress the outcomes of graduates from a given program and graduation year on the share of international students to whom first-year native students in that same program and year are exposed. For example, I regress outcomes of students who graduated from Economics at Maastricht University in 2010, such as their employment 5 years after graduation, on the share of international students among first-year Economics students at Maastricht University in 2010. The idea is that the outcomes of graduates reflect the quality of the program.

With fixed effects, none of the 18 indicators of program quality is statistically significant at conventional levels, and the coefficients are precisely estimated. These findings suggest that, conditional on program and cohort fixed effects, the share of international students does not correlate with changes in the overall quality of programs. The lack of significant differences supports the assumption that any observed effects of international students on native peers' outcomes are not driven by systematic variations in program quality.

In total, out of 114 estimates, one would expect to have 1\% of coefficients significant at the 1\% level, 5\% at the 5\% level, and 10\% at the 10\% level. The share of significant estimates in my sample is at or below the respective expectations: they are 1\%, 1\%, and 5\%, respectively.

\paragraph*{Reflection and common shocks.} Apart from the selection problem, identifying peer effects is complicated by two more issues: the reflection and common shocks problems, as outlined by \cite{Manski1993}. The reflection problem arises when international and native students potentially influence each other’s outcomes simultaneously, while common shocks occur when factors such as curricular revisions affect the entire group. These issues emerge when contemporaneous outcomes are used as treatment variables. However, in my setting, these problems are mitigated by using the share of international students as the measure of peer influence. This variable is fixed at the start of the program and is unaffected by other variables or common shocks.

\section{Results}
\label{results}

In this section, I present the main findings on the impact of exposure to international students on social ties, migration decisions, attitudes towards internationalization and migration, and labor market outcomes. Next, I conduct a heterogeneity analysis based on sex, field of study, share of females, program size, and the origin of international students. Finally, I conclude with robustness checks and a meta-analysis.

\subsection{Main results}

\paragraph*{Effect of exposure on social ties.}

I estimate the impact of studying with international students on the likelihood of living together and forming partnerships with non-native individuals 15 years after enrolling in a program. All coefficients represent a 10 percentage point increase in the share of international students in the first year of a bachelor's program.\footnote{Figures \ref{visual_1} and \ref{visual_2} plot binned relationships between the residualized share of international students and the residualized outcomes.} Column (1) of Table \ref{results_social} shows that a 10 percentage point increase in the share of international students increases the likelihood that native students will live with non-natives by 0.6 percentage points 15 years post-enrollment. This change corresponds to a 5.9\% increase relative to the sample mean of 10.2\%.

\begingroup
\setlength{\tabcolsep}{20pt} 
\begin{table}[!htb]
\centering
{\footnotesize
\begin{threeparttable}
	\caption{The effect of exposure to international students on social ties and migration decisions of native students 15 years post-enrollment}
    \label{results_social} 
	
\begin{tabular}{llll}
  \hline 
\hline \\[-2ex]  
 & (1) & (2) & (3) \\
 & Cohabited with  & Married to & Emigrated   \\ 
 & a non-native & a non-native & \\
  \hline
  \hline
International share & 0.006 & 0.001 & 0.003 \\ 
in 10\%-points  & (0.001) & (0.001) & (0.001) \\ 
Sharpened $q$-value & [0.005] & [0.042] & [0.019] \\
         \hline
  Mean & 0.102 & 0.035 & 0.065 \\ 
  Effect size & 5.9\% & 4.2\% & 4.0\% \\ 
  Cohorts & 1988-2008 & 1988-2008 & 1988-2008 \\
  N & 605,367 & 587,676 & 617,094 \\
    \hline 
\hline 
\end{tabular}

    \begin{tablenotes}
			\footnotesize
			\item \textit{Notes}: All regressions include program and cohort fixed effects. A non-native is defined as a person without Dutch nationality. Effect size represents the ratio of the estimated coefficient to the sample mean of the outcome. Sharpened $q$-values are adjusted False Discovery Rate $p$-values, calculated using the method proposed by \cite{Anderson2008}, to account for multiple hypothesis testing. Standard errors, which are clustered at the program level, are reported in parentheses. 
	\end{tablenotes}
\end{threeparttable} 
}
\end{table}
\endgroup

Native students are also more likely to marry a non-native after exposure to more international students at university. A 10 percentage point increase in the share of international students leads to a 0.1 percentage points increase in marriages between natives and non-natives (see Column (2) of Table \ref{results_social}). This translates to a 4.2\% increase of the sample mean of 3.5\%. These findings illustrate how interactions with international students can foster closer social connections between natives and non-natives, contributing to a more integrated and culturally diverse society.

One explanation for the increased likelihood of cohabiting and marrying non-natives due to exposure to international students is the larger pool of non-natives. More international students in a program increase the chance of cohabitation and marriage between native and international students from the same program. Table \ref{social_ties_robust} shows that native students are more likely to cohabit and marry non-natives outside their program and university. These findings suggest the increased likelihood of such relationships reflects deliberate choices rather than merely convenient meeting opportunities.

\begingroup
\setlength{\tabcolsep}{15pt} 
\begin{table}[!htbp]
\centering
{\footnotesize
\begin{threeparttable}
	\caption{The effect of exposure to international students on the social ties of native students outside the program}
    \label{social_ties_robust} 
	\begin{tabular}{lllll}
  \hline
  \hline \\[-2ex] 
  & \multicolumn{1}{c}{(1)} & \multicolumn{1}{c}{(2)} & \multicolumn{1}{c}{(3)} & \multicolumn{1}{c}{(4)} \\
& \multicolumn{2}{c}{Cohabited with} & \multicolumn{2}{c}{Married to} \\
& \multicolumn{2}{c}{a non-native outside the} & \multicolumn{2}{c}{a non-native outside the} \\
& program & university & program &  university \\ 
  \hline
  \hline
International share & 0.006 & 0.005 & 0.001 & 0.001 \\ 
in 10\%-points & (0.001) & (0.001) & (0.001) & (0.001) \\ 
Sharpened $q$-value & [0.005] & [0.005] & [0.042] & [0.044] \\
         \hline
  Mean  & 0.101 & 0.100 & 0.035 & 0.035 \\ 
  Effect size & 5.5\% & 4.9\% & 3.9\% & 3.8\% \\
  Cohorts & 1988-2008 & 1988-2008 & 1988-2008 & 1988-2008 \\
  N  & 605,367 & 605,367 & 587,676 & 587,676 \\ 
  \hline 
\hline 
\end{tabular}

    \begin{tablenotes}
			\footnotesize
			\item \textit{Notes}: All regressions include program and cohort fixed effects. A non-native is defined as a person without Dutch nationality. Effect size represents the ratio of the estimated coefficient to the sample mean of the outcome. Sharpened $q$-values are adjusted False Discovery Rate $p$-values, calculated using the method proposed by \cite{Anderson2008}, to account for multiple hypothesis testing. Standard errors, which are clustered at the program level, are reported in parentheses. 
	\end{tablenotes}
\end{threeparttable} 
}
\end{table}
\endgroup

\paragraph*{Effect of exposure on migration decisions.}

Studying alongside international students can broaden native students' perspectives and increase their interest in other countries. For example, it can affect the propensity of native students to seek experiences abroad. Column (3) of Table \ref{results_social} shows that a 10 percentage point increase in the share of international students increases the probability of emigration by 0.3 percentage points, which is 4\% increase of the sample mean of 6.5\%. This suggests that universities make native students more internationally oriented.

\paragraph{Effect of exposure on attitudes towards internationalization and migration.} Exposure to international students might also shift native students' preferences, making them more open to other cultures. To examine this, I analyze how international student exposure influences natives’ attitudes towards internationalization and migration using two survey datasets.

\begingroup
\setlength{\tabcolsep}{4pt}
\begin{table}[!htb]
\hspace{-1.5cm} 
{\footnotesize
\begin{threeparttable}
	\caption{The effect of exposure to international students on attitudes towards internationalization and migration of native students}
    \label{results_survey_nse} 
	\begin{tabular}{p{9.5cm} lllll}
  \hline
  \hline  
  & Coefficient & s.e. & Sharpened & Mean & N  \\ 
  & & & $q$-value &  & \\ 
  \hline
  \hline \\[-2ex] 
  \textbf{Panel A}: Attitudes towards internationalization  & & &  &   \\
   1. How satisfied are you with the level of internationalization? (1 - very dissatisfied, 5 - very satisfied) & 0.053 & (0.019) & [0.015] & 3.524 & 123,038 \\[4ex] 
   2. How satisfied are you with the level of encouragement to learn about other cultures? (1 - very dissatisfied, 5 - very satisfied) & 0.038 & (0.019) & [0.042] & 3.197 & 123,038 \\[2ex] 
       \hline
       \hline \\[-2ex]  
   \textbf{Panel B}: Attitudes towards migration & & &  &  & \\ 
   1. Do you agree that people of foreign origin who legally reside in the Netherlands should be entitled to the same social security as Dutch citizens? (1 - fully disagree, 5 - fully agree) & 0.238 & (0.095) & [0.021] & 3.790 & 1,180 \\[6ex] 
   2. Do you agree that it does not help a neighborhood if many people of foreign origin or descent move in? (1 - fully disagree, 5 - fully agree) & -0.138 & (0.062) & [0.031] & 3.258 & 1,180 \\[6ex] 
   3. Where would you place yourself on a scale from 1 to 5, where 1 means that European unification has already gone too far and 5 means that it should go further? & 0.239 & (0.068) & [0.005] & 2.148 & 1,166 \\
  \hline
  \hline
\end{tabular}

    \begin{tablenotes}
			\footnotesize
			\item \textit{Notes}: All regressions include program and cohort fixed effects. A coefficient represents a 10 percentage point increase in the share of international students. Sharpened $q$-values are adjusted False Discovery Rate $p$-values, calculated using the method proposed by \cite{Anderson2008}, to account for multiple hypothesis testing. In Panel A, outcomes are based on National Student Survey data. In Panel B, outcomes are based on Longitudinal Internet Studies for the Social Sciences Survey data. Standard errors, which are clustered at the program level, are reported in parentheses.
	\end{tablenotes}
\end{threeparttable} 
}
\end{table}
\endgroup

I find that exposure to international students positively affects students' perceptions of internationalization (see Panel A of Table \ref{results_survey_nse}).\footnote{In Panel A of Table \ref{results_nse_robust}, I check for selection bias in programs that participate in the NSS panel and their responses to internationalization-related questions. I find no correlation between the share of international students and the probability of a program being in the NSS panel or answering any of the internationalization-related questions. Panels B and C of Table \ref{results_nse_robust} show the main findings using linear regression with weights and ordered probit regression. The results are consistent with those from the linear regression in Table \ref{results_survey_nse}.} A 10 percentage point increase in the share of international students increases satisfaction with internationalization by 0.053 points (0.12 SD) and satisfaction with encouragement to learn about other cultures by 0.038 points (0.07 SD).

Exposure affects views of natives not only during university but also post-graduation (see Panel B of Table \ref{results_survey_nse}).\footnote{In Panel A of Table \ref{mechanisms_ordered_probit}, I check for selection bias of native students into the LISS panel and their responses to immigration-related questions. I find no correlation between the share of international students and the probability of being in the LISS panel or answering any of the immigration-related questions. Panels B and C of Table \ref{mechanisms_ordered_probit} present the main findings using linear regression with weights and ordered probit regression. The results are similar to those from linear regression without weights in Table \ref{results_survey_nse}.} A 10 percentage point increase in the share of international students in a university program leads to a 0.238 point increase (0.30 SD) in agreement that people of foreign origin who legally reside in the Netherlands should be entitled to the same social security as Dutch citizens. Furthermore, it leads to a 0.138 point decrease ($-0.16$ SD) in agreement that it does not help a neighborhood if many people of foreign origin or descent move in. Additionally, exposure to international students shifts opinions on European unification, with a 0.239 point increase (0.28 SD) in the belief that European unification should go further. 

The results indicate that exposure to international students leads to deliberate and lasting social ties between native and non-native students. Native students are more likely to form relationships with non-natives outside their academic environment, reflecting a preference shift. This exposure also influences migration decisions and shapes natives' views on internationalization and migration, highlighting the important role universities play in encouraging long-term social integration and openness.

\paragraph*{Effect of exposure on labor market outcomes.}

After establishing the positive effects of exposure on the social ties and openness to migration among native students, I examine whether the presence of international students impacts the labor market outcomes of natives. Policymakers often express concerns that international students might strain educational resources and negatively affect the quality of education, potentially disadvantaging native students' labor market outcomes. Table \ref{results_labor} presents the findings for employment, income percentile, entrepreneurship, and the share of foreign-born co-workers. Across all outcomes, the estimates are precisely estimated zero. The precision of the estimated coefficients allows me to rule out any effects more negative than $-0.2$ percentage points on employment, income, and entrepreneurship, as well as effects more negative than $-0.1$ percentage point on the share of foreign-born co-workers.\footnote{Emigration is not a random event. In the period prior to leaving the Netherlands, Dutch students who later emigrated were less likely to be employed, ranked lower in the income distribution, were less likely to be entrepreneurs, and tended to work in more international firms (Table \ref{emigration_missing}). These patterns suggest that emigrants were less attached to the domestic labor market. Since realized outcomes are not observed for emigrants, I estimate effects of exposure only for those who remain in the country and address missing outcomes through bounding analyses.

In the first approach, I impose extreme assumptions by assigning emigrants either very low or very high outcomes. For binary outcomes (employment and entrepreneurship), I impute zeros for lower bounds and ones for upper bounds. For continuous outcomes (income percentile and the share of foreign-born co-workers), I use values from the bottom 10\% and top 10\% of peers in the program of emigrants. Panels A and B of Table \ref{emigration_imputation} show that the resulting bounds are tight and close to zero. The only statistically significant estimate is the lower bound for employment, which suggests that exposure to a higher share of international students reduces employment by 0.3 percentage points, a small effect relative to the sample mean of 88.2\%. 

In the second approach, I make a less extreme assumption and impute outcomes of emigrants from the year before their last observed year. The results, shown in Panel C of Table \ref{emigration_imputation}, are very similar to those from the upper- and lower-bound analysis, again producing tight bounds close to zero. The only exception is employment, with an estimated effect of $-0.2$ percentage points. Overall, the bounding analyses indicate that any bias from missing outcomes of emigrants is negligible, with estimated effects remaining very small and close to zero.} Overall, the findings point to a lack of long-term impact on the labor market.

\begingroup
\setlength{\tabcolsep}{10pt} 
\begin{table}[!htb]
\centering
{\footnotesize
\begin{threeparttable}
	\caption{The effect of exposure to international students on labor market outcomes of native students 15 years post-enrollment}
    \label{results_labor} 
	
\begin{tabular}{llllll}
  \hline 
\hline \\[-2ex]  
& (1) & (2) & (3) & (4) \\
 & Employed & Income & Entrepreneur & \% of foreign-born  \\ 
& & percentile & & co-workers  \\ 
  \hline
  \hline
International share & -0.001 & 0.001 & 0.000 & 0.001  \\ 
in 10\%-points & (0.001) & (0.001) & (0.001) & (0.001) \\ 
Sharpened $q$-value & [0.077] & [0.191] & [0.257] & [0.109] \\
         \hline
  Mean & 0.943 & 0.717 & 0.072 & 0.122 \\ 
  Effect size & -0.1\% & 0.1\% & 0.1\% & 0.6\% \\ 
 Cohorts & 1988-2008 & 1988-2008 & 1988-2008 & 1991-2008 \\
  N & 576,961 & 576,961 & 576,961 & 427,687 \\ 
    \hline 
\hline 
\end{tabular}

    \begin{tablenotes}
			\footnotesize
			\item \textit{Notes}: All regressions include program and cohort fixed effects. Effect size represents the ratio of the estimated coefficient to the sample mean of the outcome. Sharpened $q$-values are adjusted False Discovery Rate $p$-values, calculated using the method proposed by \cite{Anderson2008}, to account for multiple hypothesis testing. Standard errors, which are clustered at the program level, are reported in parentheses. 
	\end{tablenotes}
\end{threeparttable} 
}
\end{table}
\endgroup

This absence of negative effects on labor market outcomes is particularly significant in light of ongoing policy debates that often focus on the potentially detrimental impacts of international student presence on the educational and labor market outcomes of natives.\footnote{Table \ref{results_education} and Panel A of Table \ref{native_flight_stem} report additional estimates of the effect of exposure to international students on educational outcomes of natives, including the probability of dropping out in the first year, switching programs within the same university, transferring to another university while remaining in the same program, time to graduation, and completion of bachelor’s and master’s degrees. In line with the labor market results, none of these outcomes are statistically significant.} The precision of these estimates refutes concerns about statistically and economically significant negative effects, which suggests that exposure to international peers does not harm native students' local economic success.

\subsection{Non-linearity} 

The effects of exposure to international students may plausibly be nonlinear. For instance, \cite{Anderberg2024} document an inverse-U shaped relationship between immigrant share and natives’ in-group bias: bias initially increases with diversity but declines once immigrants form a sufficiently large group to reshape the peer environment. Motivated by this insight, I examine whether the relationship between the share of international students and native students' outcomes similarly deviates from linearity. Following the approach of \cite{Anderberg2024}, I estimate a cubic specification of the form: 
\begin{align}
	\label{model_nonlinear}
	Y_{ifut} = \beta_1 D_{fut} + \beta_2 D^2_{fut} + \beta_3 D^3_{fut} + \alpha_{fu} + \tau_{ft} + \varepsilon_{ifut}
\end{align}
After deriving the marginal effects, I use the delta method to compute standard errors and plot the results with 95\% confidence intervals in Figures \ref{marginal_1} and \ref{marginal_2}. 

Unlike \cite{Anderberg2024}, I find limited evidence of non-linear effects. The marginal effects are consistently positive for social outcomes, such as cohabitation and marriage, indicating that greater exposure to international peers strengthens native students’ social ties across the entire distribution of exposure. For emigration, the marginal effects are significantly positive at exposure levels up to 20\%, but becomes statistically insignificant at higher levels. This suggests that even modest exposure is sufficient to increase international mobility, while additional exposure does not further reinforce this effect. In contrast, the marginal effects on labor market outcomes are not statistically significant at any level of international student presence, with the exception of income. Overall, the findings indicate that the effects of exposure to international students are stable across the distribution of international student presence, with positive effects on social outcomes and no significant effects on labor market outcomes.

\subsection{Heterogeneity}

In this section, I examine how international student exposure impacts groups with different social dynamics, competitive pressures, network diversity, and cultural similarities. These factors can influence both social integration and professional outcomes, with some environments encouraging stronger cross-cultural connections and others limiting them. By analyzing how exposure effects vary across these underlying features, I aim to identify which contexts enhance or weaken the benefits of international student exposure.

\paragraph{Sex.} Prior research suggests that men and women may experience social and professional networks differently \citep{Beaman2018, Cullen2023}. This could influence integration outcomes between native and international students by sex. To investigate this, I interact a female dummy with the share of international students. Panel A of Table \ref{results_hetero} indicates no statistically significant differences by sex, suggesting that male and female natives experience similar impacts from exposure to international students.

\paragraph{Field of study.} Social engagement patterns can vary by field of study. STEM fields, with their competitive nature and intensive coursework, may provide fewer opportunities for social integration than non-STEM fields. Panel B of Table \ref{results_hetero} shows that non-STEM graduates benefit from social integration with international students, while STEM graduates show no significant increase in the probability to cohabit with a non-native. This suggests that the competitive environment of STEM fields may limit opportunities for cross-cultural interactions.\footnote{Consistent with \cite{Anelli2023}, I find evidence of ``native flight'' from STEM programs. I find that natives in STEM programs are more likely to drop out when exposed to more international students (see Panel B of Table \ref{native_flight_stem}). This contrasts with non-STEM fields, where I find no such effect, in line with the broader pattern that positive social effects are concentrated outside STEM.}

\paragraph{Share of females.} Research suggests that men tend to be more competitive than women \citep{Niederle2007, Buser2014}. In highly male-dominated settings, competition may take precedence over collaboration, potentially limiting opportunities for social integration. To test this, I examine whether the effect of exposure to international students varies by the sex composition of a program. I use a 20\% female threshold because programs with such a low share of women are heavily male-dominated, where competitive norms are likely to be most pronounced. Panel C of Table \ref{results_hetero} shows that in programs with fewer than 20\% female students, exposure to international students has no significant effect on the probability to cohabit with a non-native. This finding suggests that the competitive nature in male-dominated environments may suppress the social integration benefits of international exposure.

\paragraph{Program size.} Larger programs may promote greater social integration, as they offer broader networks and more opportunities for students to engage with peers. Panel D of Table \ref{results_hetero} confirms this, showing that natives in larger programs experience stronger effects on social ties and migration decisions. Exposure in larger programs also increases the likelihood of employment in firms with more foreign-born co-workers, suggesting that these broader environments foster integration and may encourage interest in global career paths.

\paragraph{EEA+ and non-EEA+ students.} Cultural and institutional similarities between EEA+ students and Dutch natives may facilitate social connections, as they often share cultural and institutional backgrounds. This familiarity likely lowers barriers to interaction, resulting in stronger integration outcomes. Columns (7) and (8) of Tables \ref{sensitivity_1} and \ref{sensitivity_2} show that exposure to EEA+ students significantly boosts social integration. In contrast, exposure to non-EEA+ students, who may face greater cultural and linguistic differences, shows no significant integration effects. However, exposure to non-EEA+ peers increases satisfaction with opportunities to learn about other cultures, showing that these interactions promote cultural exchange. These results suggest that shared cultural contexts with EEA+ students enhance social ties, while interactions with non-EEA+ students contribute to broader cultural awareness, highlighting the value of diversity in the learning environment.

\subsection{Robustness}
\label{robustness}

The main analysis uses a model with program and cohort fixed effects. Here, I assess the robustness of the findings using alternative model specifications, sample definitions, treatment definitions, and a placebo test.

\paragraph*{Correction for multiple hypothesis testing.} Given the number of hypotheses tested, I apply the correction method from \cite{Anderson2008} to control for multiple hypothesis testing. This correction reduces the risk of false positives, and the results remain statistically significant at the 5\% level (see sharpened $q$-values in Tables \ref{results_social}-\ref{results_labor}).

\paragraph*{Outcome measurement at 10 and 25 years.} In the baseline analysis, I measure outcomes 15 years post-enrollment. To test for effects over shorter and longer horizons, I estimate outcomes at 10 and 25 years post-enrollment in Table \ref{results_different_periods}. Results confirm the baseline findings: increased social ties between natives and non-natives, higher emigration rates, and no significant labor market effects.

\paragraph*{Including time trends.} To control for potential confounding from time-related factors, I include linear and quadratic field specific time trends. Columns (2) and (3) in Tables \ref{robustness_1} and \ref{robustness_2} show that the results remain robust to this adjustment.

\paragraph*{Controlling for program size.} To ensure that the results are not influenced by the size of the programs, I include program size as a control variable in the model. Column (4) in Tables \ref{robustness_1} and \ref{robustness_2} shows that the results remain robust.

\paragraph*{Controlling for individual characteristics.} To account for possible confounding from individual characteristics correlated with the share of international students, I include individual controls. Column (5) in Tables \ref{robustness_1} and \ref{robustness_2} shows that the results stay robust.

\paragraph*{Controlling for peer characteristics.} Although individual characteristics do not impact estimates, the share of international students could be correlated with peer characteristics. In Column (6) in Tables \ref{robustness_1} and \ref{robustness_2}, I control for these characteristics, and the results remain unchanged.

\paragraph*{Including first-generation migrants.} The main analysis excludes first-generation migrants due to ambiguity in categorizing them as native or international. Column (8) in Tables \ref{robustness_1} and \ref{robustness_2} shows that including them does not change the results.

\paragraph*{Including older first-year students.} To focus on typical undergraduate populations, I exclude students over 30 at enrollment in the main analysis. Column (9) in Tables \ref{robustness_1} and \ref{robustness_2} shows that including them does not affect the results.

\paragraph{Sensitivity to alternative definitions of international students.} 

\cite{Angrist2014} and \cite{Feld2017} highlight the potential for measurement error to bias estimates of peer effects. The availability of comprehensive administrative data in my research allows for an examination of how different definitions of international students influence the sensitivity of my findings. Various definitions have been used in the literature to identify international students, including:
\begin{enumerate}
\item Having a different nationality \citep{Borjas2004, Borjas2006, Shen2016, Chevalier2020, Anelli2023, Beine2023, Rakesh2023, Chen2023, Zhu2023}.
\item Being born in a different country \citep{Machin2017, Shih2017}.
\item Residing in a different country before university enrollment \citep{Costas2023}.
\item Completing secondary education in a different country.
\end{enumerate}
These variations in definitions pose two main challenges. Firstly, they complicate the comparison of findings across studies, as it becomes unclear whether observed differences in peer effects are due to genuine variations or merely result from differing definitions. Secondly, definitions that inaccurately characterize international students can lead to biased estimates of peer effects.

To estimate the extent of potential biases, I propose a definition of a ``true" international student as someone who is both a non-national and did not complete secondary education in the host country. This definition aims to capture individuals who lack prior exposure to the host country's education system. 

Tables \ref{sensitivity_1} and \ref{sensitivity_2} show that the results remain largely consistent across alternative definitions. On average, these definitions yield similar outcomes, suggesting that previous studies using different definitions produce unbiased results, and that differences in estimated peer effects across settings are due to genuine variations.

\paragraph{Placebo analysis.}

In this section, I present the results of the placebo analysis. Instead of using the actual share of international students that native students are exposed to, I assign them a share of international students from the same program and year but a different university.\footnote{Programs that are only offered at one university in a given year are excluded from the estimation.} For instance, a student enrolled in Economics at the University of Amsterdam in 2010 is assigned the share of international students from Economics at Erasmus University in 2010. This process is repeated 1,000 times, each time estimating the effect of this placebo treatment.

The resulting normal-looking distributions of simulated counterfactual effects confirm that the true estimates are located in the tails of the distributions, further validating the robustness of the findings. Figures \ref{placebo_1}-\ref{placebo_3} display plots with estimates from 1,000 simulated datasets, the mean estimated coefficients of the randomly assigned shares, and the actual estimated coefficients. The mean estimated effect of the randomly assigned shares is centered around zero for all outcomes. In contrast, the actual estimated coefficients lie in the tails of the distribution, providing further support for the validity of the identification assumption.

\subsection{Discussion} 
\label{discussion}

To contextualize my estimates, I compare the findings of this study with two bodies of literature: \textit{(i)} the spillover effects of international students on native peers and \textit{(ii)} the effects of interacting with minority groups on prejudice.

\begin{figure}[!htb]
    \caption{Comparison with estimates from the international students literature}
    \includegraphics[width=\textwidth]{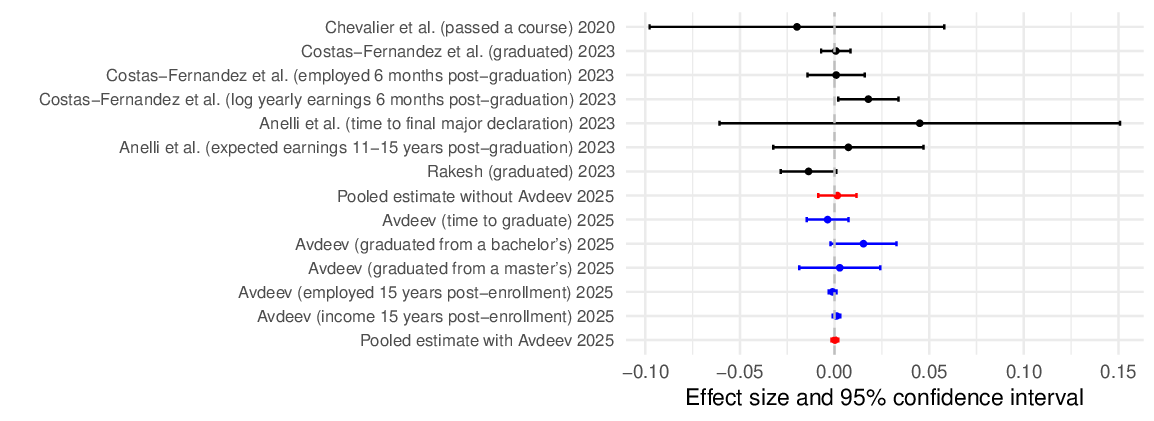}
    \label{estimates_international} \\
    {\footnotesize \textit{Notes:} This graph presents estimates from 4 studies (black) by \cite{Chevalier2020, Costas2023, Anelli2023, Rakesh2023}, 2 pooled random effects estimates calculated by the author (red), and estimates from the current paper (blue). One pooled estimate is based on the estimates from previous studies, while the second one additionally includes the estimates from the current paper. All estimates are converted to represent a 10 percentage point increase in the share of international students. Effect size represents the ratio of the estimated coefficient to the sample mean of the outcome. The estimate from \cite{Chevalier2020} is reversed from ``failed a course" to ``passed a course" to ensure comparability. Table \ref{literature} provides an overview of the estimates from previous research. Table \ref{results_education} shows the estimates of the current paper on educational attainment.}
\end{figure}

\paragraph{Peer effects of international students.} I compare my estimates with the existing literature on the peer effects of international students. Figure \ref{estimates_international} summarizes estimates from previous studies \citep{Chevalier2020, Anelli2023, Costas2023, Rakesh2023}. Since no meta-analytical study exists, I calculate a pooled random effects meta-analytic estimate for comparison. 

While previous studies report small or no effects of international students on the educational outcomes of native students, they do not establish precise nulls. The pooled meta-analytic estimate, excluding my estimates, is 0.1\% with a 95\% confidence interval $[-0.009, 0.012]$. My estimates are more precise, allowing me to rule out even very small positive and negative effects. Moreover, I extend the analysis to long-term labor market outcomes, which have been previously unexplored. When my estimates are incorporated into the meta-analysis, the pooled estimate is 0.0\% with a tighter 95\% confidence interval $[-0.002, 0.002]$. This provides evidence that studying with international students has not only no statistically significant impact but also no economically meaningful effect on the educational and labor market outcomes of native students.

\begin{figure}[!htb]
    \caption{Comparison with estimates from the contact hypothesis literature}
    \includegraphics[width=\textwidth]{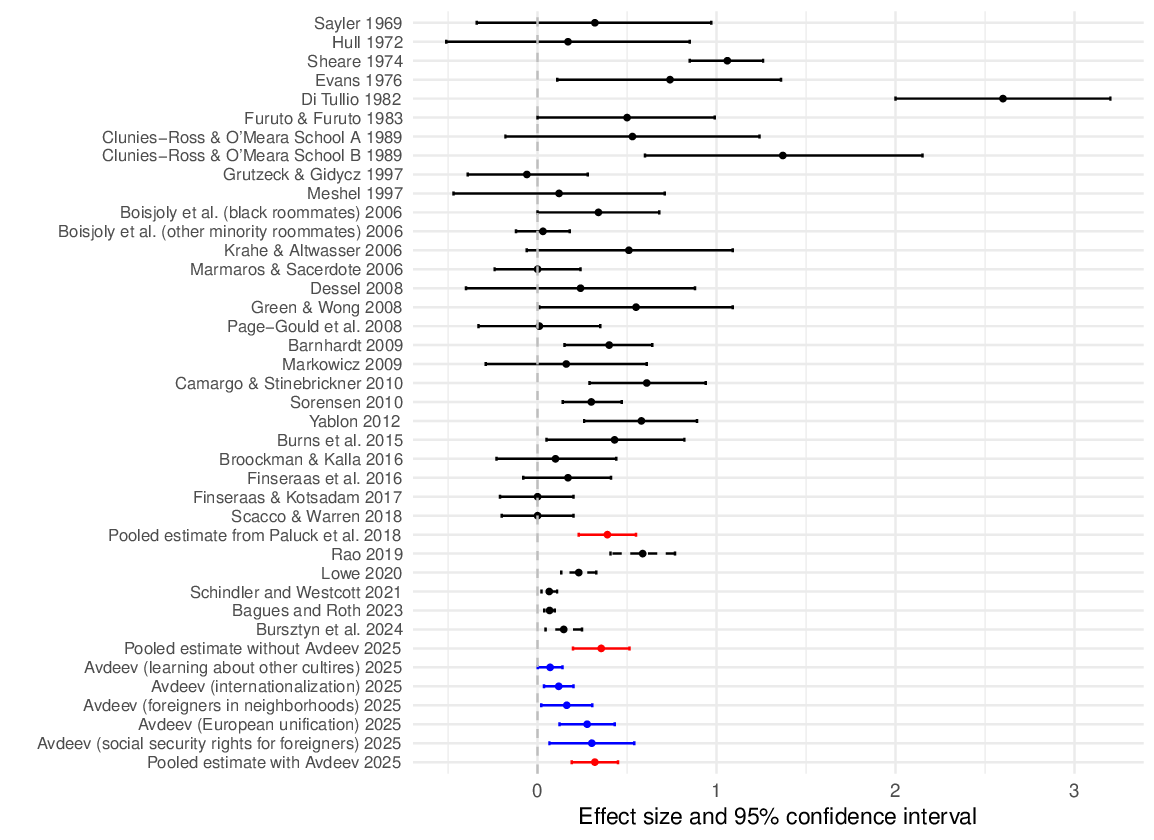}
    \label{estimates_contact} \\
    {\footnotesize \textit{Notes:} This graph presents estimates from 27 studies (black) included in a meta-analysis by \cite{Paluck2019}, 3 pooled random effects estimates (red), estimates from \cite{Rao2019}, \cite{Lowe2021}, \cite{Schindler2021}, \cite{Bagues2023}, and \cite{Bursztyn2024} (dashed), and estimates from the current paper (blue). One pooled estimate is from the meta-analysis by \cite{Paluck2019}, while the second one additionally includes the estimates from \cite{Rao2019}, \cite{Lowe2021}, \cite{Schindler2021}, \cite{Bagues2023}, \cite{Bursztyn2024}, and the third one additionally includes estimates from the current paper. \cite{Paluck2019} includes studies that randomly assigned intergroup contact and measured outcomes more than a day after treatment. Effect size represents the ratio of the estimated coefficient to the standard deviation of the outcome. The estimate from \cite{Schindler2021} and the estimate for the effect of exposure on attitudes towards foreigners in a neighborhood (foreigners in neighborhoods) are reversed so that a positive coefficient indicates an improved attitude to ensure comparability.}
\end{figure}

\paragraph{Interaction with minority groups on prejudice.} I then compare my estimates on the effects of exposure to a higher share of international students on attitudes towards internationalization and migration with studies examining similar outcomes. For instance, \cite{Paluck2019} provides a meta-analysis of the effects of intergroup contact with various minority groups (ethnic, racial, etc.) on prejudice of the majority group. \cite{Paluck2019} estimates a pooled effect size of $0.39$ SD with a 95\% confidence interval $[0.23, 0.55]$. 

My three estimates of the effect on attitudes towards migration are 0.16 SD, 0.28 SD, and 0.30 SD in absolute terms. The latter two fall within the confidence interval of the pooled effect size from \cite{Paluck2019}, suggesting that exposure to international students has a comparable impact to interventions studied in the contact hypothesis literature. This finding is notable because, unlike direct interventions, such as structured intergroup programs or assignment of roommates from different backgrounds, exposure to international students occurs more passively through everyday interactions. The fact that even this less intensive form of contact produces similar effects suggests that diversity in regular social environments can meaningfully shape attitudes towards migration.

\section{Conclusion} 
\label{conclusion}

This paper provides the first evidence on the long-term impact of international students on native students. Using three decades of Dutch survey and administrative data and exploiting variation in the share of international students within university programs over time, this study identifies several key findings. A higher share of international students in the first year of a bachelor's program strengthens native students' connections with non-natives, improves attitudes towards migration and cultural learning, and increases openness to migration. Importantly, exposure has no negative effects on labor market outcomes.

This study contributes to the literature on higher education and the contact hypothesis by focusing on the understudied group of international students. The findings underscore the dual importance of social and human capital considerations in shaping higher education funding and immigration policies. A better understanding of these relationships can help educational leaders and policymakers make informed decisions that enhance educational outcomes while fostering a more inclusive society. Policies that restrict international student enrollment may, therefore, limit broader societal benefits beyond tuition revenue.

\pagebreak

\singlespacing
\bibliographystyle{chicago}
\bibliography{International_students.bib}

\pagebreak

\section*{Supplemental Appendix}

\subsection*{University as a Melting Pot: Long-term Effects of Internationalization}

\subsubsection*{Stanislav Avdeev}

\section*{Online Appendix A: Variable Construction}
\label{variable_construction}

\setcounter{table}{0}
\renewcommand{\thetable}{A\arabic{table}}
\setcounter{figure}{0}
\renewcommand{\thefigure}{A\arabic{figure}}

In this section, I describe how I construct the variables.

\paragraph{Cohabitation.} Each year, I check whether a person is registered as living with someone at the same address. I then filter for non-native individuals at the same address. The variable equals 1 if the person has ever been registered as living with a non-native person, and 0 otherwise.

\paragraph{Marriage.} Each year, I check whether a person is married or in a registered partnership. I then select only non-native spouses. If a person has more than one marriage or registered partnership in a year, I retain the latest record. The variable equals 1 if the person has ever been married or in a registered partnership with a non-native person, and 0 otherwise.

\paragraph{Emigration.} This variable equals 1 if a person is no longer registered as living in the Netherlands and has not been reported as deceased, and 0 otherwise.

\paragraph{\% of foreign-born co-workers.} This variable measures the share of employees within a firm in the Netherlands who were born abroad.

\paragraph{Employed.} This variable equals 1 if a person is employed and 0 otherwise.

\paragraph{Income percentile.} For each individual, I calculate their income percentile within their same-age group for a given year, separately for men and women. This variable ranges between 0 and 1.

\paragraph{Entrepreneur.} This variable equals 1 if a person is an entrepreneur and 0 otherwise.

\paragraph{LISS survey questions.} Since LISS is a panel survey, I calculate the average response for each individual to a given question over years.

\paragraph{NSS survey questions.} For the NSS survey, which provides only aggregated responses, I assign each student the average response to a given question.

\pagebreak

\section*{Online Appendix B: Tables}

\begingroup
\setlength{\tabcolsep}{6pt} 
\begin{table}[!htbp]
\hspace{-2cm} 
\vspace{-2cm} 
{\footnotesize
\begin{threeparttable}
	\caption{Balancing table of characteristics of first-year students, their siblings, and parents}
    \label{balance_1} 
	\begin{tabular}{l S[table-format=-1.3] l S[table-format=-1.3] l S[table-format=1.2]}   \hline
  \hline  
  & \multicolumn{2}{l}{\textbf{No fixed effects}} & \multicolumn{3}{l}{\textbf{With fixed effects}} \\ 
  & {Coefficient} & {s.e.} & {Coefficient} & {s.e.} & {Mean} \\ 
  \hline
  \hline \\[-2ex] 
  \textbf{Panel A}: First-year students & &  & \\ 
  Female & -0.006 & (0.009) & 0.000 & (0.002) & 0.50 \\ 
  Age at enrollment & -0.109 & (0.022) & -0.018 & (0.036) & 19.78 \\ 
  Gap year & -0.011 & (0.005) & -0.001 & (0.007) & 0.33 \\ 
  Second generation immigrant & 0.006 & (0.003) & -0.002 & (0.001) & 0.14 \\ 
  Family size & -0.028 & (0.003) & -0.008 & (0.002) & 2.53 \\
  Secondary school exam grade & 0.011 & (0.013) & 0.005 & (0.005) & 0.14 \\ 
  Has a vocational degree & -0.005 & (0.003) & -0.004 & (0.005) & 0.09 \\ 
  \hline \\[-2ex] 
  \textbf{Panel B}: Siblings &  &  & \\ 
  Female & 0.001 & (0.001) & 0.001 & (0.001) & 0.49 \\ 
  Age at enrollment of the sibling & -0.133 & (0.028) & -0.020 & (0.040) & 18.54 \\ 
  Has a university degree & 0.002 & (0.001) & -0.003 & (0.002) & 0.08 \\ 
  \hline \\[-2ex] 
  \textbf{Panel C}: Parents &&&  &  & \\ 
  Father's age at enrollment of the child & 0.281 & (0.046) & 0.025 & (0.024) & 50.12  \\ 
  Mother's age at enrollment of the child & 0.274 & (0.043) & 0.028 & (0.024) & 47.83  \\ 
  Father is married or cohabiting & -0.004 & (0.001) & 0.000 & (0.001) & 0.90 \\ 
  Mother is married or cohabiting & -0.003 & (0.001) & -0.001 & (0.001) & 0.87 \\ 
  Father has a university degree & 0.031 & (0.004) & 0.002 & (0.002) & 0.18 \\ 
  Mother has a university degree & 0.030 & (0.004) & 0.002 & (0.002) & 0.17 \\  
  Father is employed & 0.001 & (0.001) & 0.001 & (0.001) & 0.93 \\ 
  Mother is employed & 0.004 & (0.001) & -0.001 & (0.001) & 0.83 \\ 
  Father is an entrepreneur & 0.003 & (0.001) & 0.001 & (0.001) & 0.11 \\ 
  Mother is an entrepreneur & 0.005 & (0.001) & 0.000 & (0.001) & 0.08 \\ 
  Father receives unemployment benefits & 0.000 & (0.000) & 0.000 & (0.000) & 0.01 \\ 
  Mother receives unemployment benefits & 0.001 & (0.000) & 0.000 & (0.000) & 0.01 \\ 
  Father receives social benefits & 0.000 & (0.001) & -0.001 & (0.001) & 0.07 \\ 
  Mother receives social benefits & 0.000 & (0.001) & 0.001 & (0.001) & 0.07 \\ 
  Father's income percentile & 0.002 & (0.002) & 0.000 & (0.001) & 0.67 \\ 
  Mother's income percentile & 0.001 & (0.001) & -0.001 & (0.001) & 0.57 \\ 
  At least one parent has a university degree & 0.042 & (0.006) & 0.004 & (0.003) & 0.26 \\
  At least one parent is employed & 0.001 & (0.001) & 0.000 & (0.001) & 0.97 \\ 
  At least one parent is self-employed & 0.008 & (0.002) & 0.001 & (0.001) & 0.18 \\ 
  At least one parent receives unemployment benefits & 0.001 & (0.000) & 0.000 & (0.001) & 0.02 \\ 
  At least one parent receives social benefits & 0.000 & (0.001) & 0.000 & (0.001) & 0.12 \\ 
  Average parents' income percentile & 0.001 & (0.001) & -0.001 & (0.001) & 0.62 \\ 
  \hline
  $F$-test and its $p$-value & \multicolumn{2}{c}{4.44 ($<0.001$)} & \multicolumn{2}{c}{1.11 (0.31)} \\ 
  N & \multicolumn{2}{c}{1,037,347} & \multicolumn{2}{c}{1,037,347} \\ 
  \hline
  \hline
\end{tabular}

    \begin{tablenotes}
			\footnotesize
			\item \textit{Notes}: Regressions with fixed effects include program and cohort fixed effects. For siblings, all outcomes are measured one year before their first sibling enrolls in a university. For parents, all outcomes are measured one year before their first child enrolls in a university. N represents the number of first-year students. Standard errors (s.e.), which are clustered at the program level, are reported in parentheses. 
	\end{tablenotes}
\end{threeparttable} 
}
\end{table}
\endgroup

\pagebreak

\begingroup
\setlength{\tabcolsep}{6pt} 
\begin{table}[!htbp]
\hspace{-2cm} 
{\footnotesize
\begin{threeparttable}
	\caption{Balancing table of characteristics of first-year students and the lagged share of international students}
    \label{balance_2} 
	\begin{tabular}{l S[table-format=-1.3] l S[table-format=-1.3] l S[table-format=1.2]}   \hline
  \hline  
  & \multicolumn{2}{l}{\textbf{No fixed effects}} & \multicolumn{3}{l}{\textbf{With fixed effects}} \\ 
  & {Coefficient} & {s.e.} & {Coefficient} & {s.e.} & {Mean} \\ 
  \hline
  \hline \\[-2ex] 
  \textbf{Panel A}: First-year students & &  & \\ 
  Female & -0.007 & (0.009) & -0.001 & (0.002) & 0.50 \\ 
  Age at enrollment & -0.105 & (0.023) & -0.007 & (0.035) & 19.78 \\ 
  Gap year & -0.010 & (0.005) & 0.000 & (0.007) & 0.33 \\ 
  Second generation immigrant & 0.005 & (0.003) & -0.003 & (0.002) & 0.14 \\ 
  Family size & -0.026 & (0.003) & -0.005 & (0.003) & 2.53 \\
  Secondary school exam grade & 0.011 & (0.013) & 0.009 & (0.005) & 0.14 \\ 
  Has a vocational degree & -0.006 & (0.003) & -0.002 & (0.005) & 0.09 \\ 
  \hline \\[-2ex] 
  \textbf{Panel B}: Siblings &  &  & \\ 
  Female & 0.002 & (0.001) & 0.002 & (0.001) & 0.49 \\ 
  Age at enrollment of the sibling & -0.125 & (0.030) & -0.002 & (0.037) & 18.54 \\ 
  Has a university degree & 0.002 & (0.002) & -0.002 & (0.002) & 0.08 \\ 
  \hline \\[-2ex] 
  \textbf{Panel C}: Parents &&&  &  & \\ 
  Father's age at enrollment of the child & 0.280 & (0.048) & 0.032 & (0.023) & 50.12  \\ 
  Mother's age at enrollment of the child & 0.274 & (0.045) & 0.033 & (0.023) & 47.83  \\ 
  Father is married  or cohabiting & -0.004 & (0.001) & -0.001 & (0.001) & 0.90 \\ 
  Mother is married  or cohabiting & -0.003 & (0.002) & 0.000 & (0.001) & 0.87 \\ 
  Father has a university degree & 0.030 & (0.004) & 0.003 & (0.002) & 0.18 \\ 
  Mother has a university degree & 0.029 & (0.005) & 0.003 & (0.002) & 0.17 \\  
  Father is employed & 0.001 & (0.001) & 0.001 & (0.001) & 0.93 \\ 
  Mother is employed & 0.004 & (0.001) & 0.000 & (0.001) & 0.83 \\ 
  Father is an entrepreneur & 0.003 & (0.001) & 0.000 & (0.001) & 0.11 \\ 
  Mother is an entrepreneur & 0.005 & (0.001) & 0.001 & (0.001) & 0.08 \\ 
  Father receives unemployment benefits & 0.000 & (0.000) & 0.000 & (0.000) & 0.01 \\ 
  Mother receives unemployment benefits & 0.000 & (0.000) & 0.000 & (0.000) & 0.01 \\ 
  Father receives social benefits & 0.000 & (0.001) & -0.001 & (0.001) & 0.07 \\ 
  Mother receives social benefits & 0.000 & (0.001) & -0.001 & (0.001) & 0.07 \\ 
  Father's income percentile & 0.002 & (0.002) & 0.001 & (0.001) & 0.67 \\ 
  Mother's income percentile & 0.001 & (0.001) & -0.001 & (0.001) & 0.57 \\ 
  At least one parent has a university degree & 0.041 & (0.006) & 0.004 & (0.003) & 0.26 \\
  At least one parent is employed & 0.001 & (0.001) & 0.000 & (0.001) & 0.97 \\ 
  At least one parent is self-employed & 0.008 & (0.002) & 0.002 & (0.001) & 0.18 \\ 
  At least one parent receives unemployment benefits & 0.001 & (0.000) & 0.000 & (0.001) & 0.02 \\ 
  At least one parent receives social benefits & 0.000 & (0.001) & -0.001 & (0.001) & 0.12 \\ 
  Average parents' income percentile & 0.002 & (0.001) & 0.000 & (0.001) & 0.62 \\ 
  \hline
  $F$-test and its $p$-value & \multicolumn{2}{c}{3.90 ($<0.001$)} & \multicolumn{2}{c}{1.30 (0.12)} \\ 
  N & \multicolumn{2}{c}{997,173} & \multicolumn{2}{c}{997,173} \\ 
  \hline
  \hline
\end{tabular}

    \begin{tablenotes}
			\footnotesize
			\item \textit{Notes}: Regressions with fixed effects include program and cohort fixed effects. For siblings, all outcomes are measured one year before their first sibling enrolls in a university. For parents, all outcomes are measured one year before their first child enrolls in a university. N represents the number of first-year students for whom information on the share of international students in the previous year is available. Standard errors (s.e.), which are clustered at the program level, are reported in parentheses. 
	\end{tablenotes}
\end{threeparttable} 
}
\end{table}
\endgroup

\pagebreak

\begingroup
\setlength{\tabcolsep}{6pt} 
\begin{table}[!htbp]
\hspace{-2cm} 
{\footnotesize
\begin{threeparttable}
	\caption{Balancing table of peer characteristics of first-year students}
    \label{balance_3} 
	\begin{tabular}{l S[table-format=-1.3] l S[table-format=-1.3] l S[table-format=1.2]}
  \hline
  \hline  
  & \multicolumn{2}{l}{\textbf{No fixed effects}} & \multicolumn{3}{l}{\textbf{With fixed effects}} \\ 
  & {Coefficient} & {s.e.} & {Coefficient} & {s.e.} & {Mean} \\ 
  \hline
  \hline \\[-2ex] 
  \textbf{Panel A}: Peers &  &  & \\ 
  Share of females & 0.007 & (0.006) & 0.002 & (0.003) & 0.52 \\ 
  Average age at enrollment & -0.081 & (0.024) & 0.045 & (0.036) & 20.04 \\ 
  Share with a gap year & -0.007 & (0.005) & 0.010 & (0.006) & 0.38 \\ 
  Share of second generation immigrants & 0.007 & (0.003) & 0.002 & (0.003) & 0.14 \\ 
  Average family size & -0.038 & (0.007) & -0.009 & (0.006) & 2.54 \\ 
  Average secondary school exam grade & 0.010 & (0.009) & 0.011 & (0.007) & 0.20 \\ 
  Share with a vocational degree & -0.007 & (0.002) & 0.001 & (0.004) & 0.08 \\ 
  \hline \\[-2ex] 
  \textbf{Panel B}: Siblings of peers &&&  &  & \\ 
  Share of female siblings & 0.001 & (0.001) & 0.000 & (0.002) & 0.49 \\ 
  Average age at enrollment of the sibling  & -0.111 & (0.024) & 0.042 & (0.046) & 18.77 \\ 
  Share of siblings with a university degree  & 0.001 & (0.001) & -0.001 & (0.002) & 0.08 \\ 
  \hline \\[-2ex] 
  \textbf{Panel C}: Parents of peers &&&  &  & \\ 
  Average father's age at enrollment of the child  & 0.214 & (0.031) & 0.061 & (0.040) & 50.32 \\ 
  Average mother's age at enrollment of the child  & 0.198 & (0.024) & 0.049 & (0.034) & 48.00 \\ 
  Share of married  or cohabiting fathers & -0.004 & (0.001) & 0.000 & (0.002) & 0.90 \\  
  Share of married  or cohabiting mothers  & -0.003 & (0.001) & -0.002 & (0.002) & 0.87 \\ 
  Share of fathers with a university degree  & 0.026 & (0.002) & 0.001 & (0.002) & 0.16 \\ 
  Share of mothers with a university degree  & 0.025 & (0.003) & -0.001 & (0.002) & 0.15 \\ 
  Share of employed fathers & 0.000 & (0.001) & 0.000 & (0.002) & 0.93 \\ 
  Share of employed mothers & 0.005 & (0.001) & 0.000 & (0.002) & 0.82 \\ 
  Share of fathers who are entrepreneurs & 0.004 & (0.001) & 0.003 & (0.003) & 0.11 \\ 
  Share of mothers who are entrepreneurs  & 0.005 & (0.001) & 0.001 & (0.002) & 0.08 \\ 
  Share of fathers who receive unemployment benefits & 0.000 & (0.000) & 0.000 & (0.001) & 0.01 \\ 
  Share of mothers who receive unemployment benefits  & 0.000 & (0.000) & 0.001 & (0.001) & 0.01 \\ 
  Share of fathers who receive social benefits & 0.001 & (0.001) & 0.000 & (0.002) & 0.07 \\ 
  Share of mothers who receive social benefits  & 0.001 & (0.001) & 0.003  & (0.002) & 0.07 \\ 
  Average father's income percentile  & 0.001 & (0.001) & 0.000 & (0.002) & 0.65 \\ 
  Average mother's income percentile  & 0.003 & (0.001) & 0.000 & (0.002) & 0.56 \\ 
  Share of parents with a university degree & 0.035 & (0.003) & 0.000 & (0.002) & 0.23 \\ 
  Share of employed parents & 0.000 & (0.001) & -0.001 & (0.001) & 0.96 \\ 
  Share of parents who are self-employed & 0.007 & (0.001) & 0.003 & (0.003) & 0.17 \\ 
  Share of parents who receive unemployment benefits & 0.001 & (0.000) & 0.002 & (0.001) & 0.02 \\ 
  Share of parents who receive social benefits  & 0.002 & (0.001) & 0.002 & (0.002) & 0.12 \\ 
  Average parents' income percentile  & 0.002 & (0.001) & 0.000  & (0.001) & 0.60 \\ 
  \hline
  $F$-test and its $p$-value & \multicolumn{2}{c}{4.35 ($<0.001$)} & \multicolumn{2}{c}{1.03 (0.42)} \\ 
  N & \multicolumn{2}{c}{11,449} & \multicolumn{2}{c}{11,449} \\ 
  \hline
  \hline
\end{tabular}

    \begin{tablenotes}
			\footnotesize
			\item \textit{Notes}: Regressions with fixed effects include program and cohort fixed effects. For siblings, all outcomes are measured one year before their first sibling enrolls in a university. For parents, all outcomes are measured one year before their first child enrolls in a university. N represents the number of programs in which first-year students are enrolled. Standard errors (s.e.), which are clustered at the program level, are reported in parentheses. 
	\end{tablenotes}
\end{threeparttable} 
}
\end{table}
\endgroup

\pagebreak

\begingroup
\setlength{\tabcolsep}{6pt} 
\begin{table}[!htb]
\hspace{-2cm} 
{\footnotesize
\begin{threeparttable}
	\caption{Balancing table of indicators of program quality}
    \label{balance_4} 
	\begin{tabular}{l S[table-format=-1.3] l S[table-format=-1.3] l S[table-format=1.2]}
  \hline
  \hline  
  & \multicolumn{2}{l}{\textbf{No fixed effects}} & \multicolumn{3}{l}{\textbf{With fixed effects}} \\ 
  & {Coefficient} & {s.e.} & {Coefficient} & {s.e.} & {Mean} \\ 
  \hline
  \hline \\[-2ex] 
  \textbf{Panel A}: Educational outcomes &&&  &  & \\ 
  Program size & 17.755 & (4.712) & 5.624 & (4.037) & 136.76 \\ 
  Share of ever dropped out & 0.006 & (0.003) & 0.002 & (0.004) & 0.24  \\ 
  Average time to graduate & -0.075 & (0.018) & 0.017 & (0.020) & 4.65  \\ 
  Share who completed a master's degree & 0.044 & (0.004) & -0.001 & (0.004) & 0.41 \\ 
  \hline \\[-2ex] 
  \textbf{Panel B}: Social outcomes &&&  &  & \\ 
  Share who cohabited with a non-native 5 year post-graduation  & 0.012 & (0.002) & 0.000 & (0.002) & 0.08 \\
  Share who cohabited with a non-native 10 years post-graduation & 0.012 & (0.002) & 0.000 & (0.003) & 0.11 \\ 
  Share who married to a non-native 5 years post-graduation & 0.000 & (0.000) & 0.000 & (0.001) & 0.02 \\
  Share who married to a non-native 10 years post-graduation & 0.003 & (0.001) & -0.002 & (0.002) & 0.04 \\
  Emigrated 5 years post-graduation & 0.007 & (0.001) & 0.000 & (0.001) & 0.05 \\  
  Emigrated 10 years post-graduation & 0.012 & (0.002) & 0.002 & (0.003) & 0.07 \\ 
  \hline \\[-2ex] 
  \textbf{Panel C}: Labor market outcomes &&&  &  & \\ 
  Share of employed 5 year post-graduation & -0.001 & (0.001) & 0.000 & (0.002) & 0.95 \\ 
  Share of employed 10 years post-graduation & -0.001 & (0.001) & -0.001 & (0.002) & 0.95 \\ 
  Average income percentile 5 year post-graduation & 0.003 & (0.002) & -0.001 & (0.002) & 0.67 \\ 
  Average income percentile 10 years post-graduation & 0.001 & (0.003) & -0.002 & (0.002) & 0.71 \\ 
  Share of entrepreneurs 5 year post-graduation & -0.003 & (0.001) & -0.001 & (0.001) & 0.04 \\
  Share of entrepreneurs 10 years post-graduation & -0.004 & (0.001) & -0.001 & (0.002) & 0.07 \\
  \% of foreign-born co-workers 5 year post-graduation & 0.004 & (0.001) & -0.001 & (0.001) & 0.13 \\
  \% of foreign-born co-workers 10 years post-graduation & 0.003 & (0.001) & 0.001 & (0.002) & 0.13 \\
   \hline
  $F$-test and its $p$-value & \multicolumn{2}{c}{2.04 (0.006)} & \multicolumn{2}{c}{0.52 (0.95)} \\ 
  N & \multicolumn{2}{c}{10,329} & \multicolumn{2}{c}{10,329} \\ 
  \hline
  \hline
\end{tabular}

    \begin{tablenotes}
			\footnotesize
			\item \textit{Notes}: Regressions with fixed effects include program and cohort fixed effects. All outcomes, except for program size, are measured among graduates of the programs in the same year the share of international students among first-year students is measured. A non-native is defined as a person without Dutch nationality. N represents the number of programs from which students have graduated. Standard errors (s.e.), which are clustered at the program level, are reported in parentheses. 
	\end{tablenotes}
\end{threeparttable} 
}
\end{table}
\endgroup

\pagebreak

\begin{table}[!htbp]
\hspace{-1.5cm} 
{\footnotesize
\begin{threeparttable}
	\caption{The effect of exposure to international students on attitudes towards internationalization}
    \label{results_nse_robust} 
	\begin{tabular}{p{11cm} ll l l}
  \hline
  \hline  
  & Coefficient & s.e. & Mean & N  \\ 
  \hline
  \hline \\[-2ex] 
  \textbf{Panel A}: Selection into the NSS panel & & & & \\
   1. Is a program observed in the NSS panel? (0 - no, 1 - yes) & -0.007 & (0.011) & 0.911 & 381,671 \\ [1ex]
   2. How satisfied are you with the level of internationalization? (0 - not answered, 1 - answered) & 0.000 & (0.000) & 0.354 & 347,615 \\[4ex] 
   3. How satisfied are you with the level of encouragement to learn about other cultures? (0 - not answered, 1 - answered) & 0.000 & (0.000) & 0.354 & 347,615 \\[1ex]
    \hline
  \hline \\[-2ex] 
  \textbf{Panel B}: Linear regression with weights & & & & \\
   1. How satisfied are you with the level of internationalization? (1 - very dissatisfied, 5 - very satisfied) & 0.060 & (0.019) & 3.524 & 123,038 \\[4ex] 
   2. How satisfied are you with the level of encouragement to learn about other cultures? (1 - very dissatisfied, 5 - very satisfied) & 0.046 & (0.017) & 3.197 & 123,038 \\[2ex] 
  \hline
  \hline \\[-2ex] 
  \textbf{Panel C}: Ordered probit regression & & & & \\
   1. How satisfied are you with the level of internationalization? (1 - very dissatisfied, 5 - very satisfied) & 0.156 & (0.011) & 3.524 & 123,038 \\[4ex] 
   2. How satisfied are you with the level of encouragement to learn about other cultures? (1 - very dissatisfied, 5 - very satisfied) & 0.068 & (0.010) & 3.197 & 123,038 \\[2ex] 
  \hline
  \hline
\end{tabular}
    \begin{tablenotes}
			\footnotesize
			\item \textit{Notes}: A coefficient represents a 10 percentage point increase in the share of international students. In Panels A and B, all regressions include program and cohort fixed effects. In Panel C, all regressions include the share of international students after residualizing it for program and cohort fixed effects. In Panel B, I weigh observations by the ratio of students who answered the survey to the total number of students enrolled in the program. Standard errors, which are clustered at the program level, are reported in parentheses. 
	\end{tablenotes}
\end{threeparttable} 
}
\end{table}

\pagebreak

\begin{table}[!htbp]
\hspace{-2.5cm} 
{\footnotesize
\begin{threeparttable}
	\caption{The effect of exposure to international students on attitudes towards migration}
    \label{mechanisms_ordered_probit} 
	\begin{tabular}{p{13cm} ll l l}
  \hline
  \hline  
  & Coefficient & s.e. & Mean & N  \\ 
  \hline
  \hline \\[-2ex] 
  \textbf{Panel A}: Selection into the LISS panel and answering a question & & & & \\
   1. Is a person observed in the LISS panel? (0 - no, 1 - yes) & 0.000 & (0.000) & 0.001 & 1,037,347 \\ [1ex] 
   2. Do you agree that people of foreign origin who legally reside in the Netherlands should be entitled to the same social security as Dutch citizens? (0 - not answered, 1 - answered) & -0.047 & (0.036) & 0.821 & 1,437 \\ [6ex] 
   3. Do you agree that it does not help a neighborhood if many people of foreign origin or descent move in? (0 - not answered, 1 - answered) & -0.047 & (0.036) & 0.821 & 1,437 \\ [4ex] 
   4. Where would you place yourself on a scale from 1 to 5, where 1 means that European unification has already gone too far and 5 means that it should go further? (0 - not answered, 1 - answered) & -0.045 & (0.036) & 0.811 & 1,437 \\ [6ex]
   \hline
  \hline \\[-2ex] 
  \textbf{Panel B}: Linear regression with weights & & & & \\
   1. Do you agree that people of foreign origin who legally reside in the Netherlands should be entitled to the same social security as Dutch citizens? (1 - fully disagree, 5 - fully agree) & 0.282 & (0.128) & 3.790 & 1,180 \\[6ex] 
   2. Do you agree that it does not help a neighborhood if many people of foreign origin or descent move in? (1 - fully disagree, 5 - fully agree) & -0.157 & (0.062) & 3.258 & 1,180 \\[4ex] 
   3. Where would you place yourself on a scale from 1 to 5, where 1 means that European unification has already gone too far and 5 means that it should go further? & 0.162 & (0.082) & 2.148 & 1,166 \\[2ex]
  \hline
  \hline \\[-2ex] 
  \textbf{Panel C}: Ordered probit regression & & & & \\
   1. Do you agree that people of foreign origin who legally reside in the Netherlands should be entitled to the same social security as Dutch citizens? (1 - fully disagree, 5 - fully agree) & 0.296 & (0.104) & 3.790 & 1,180 \\[6ex] 
   2. Do you agree that it does not help a neighborhood if many people of foreign origin or descent move in? (1 - fully disagree, 5 - fully agree) & -0.187 & (0.102) & 3.258 & 1,180 \\[4ex] 
   3. Where would you place yourself on a scale from 1 to 5, where 1 means that European unification has already gone too far and 5 means that it should go further? & 0.253 & (0.102) & 2.148 & 1,166 \\[2ex]
  \hline
  \hline
\end{tabular}
    \begin{tablenotes}
			\footnotesize
			\item \textit{Notes}: A coefficient represents a 10 percentage point increase in the share of international students. In Panels A and B, all regressions include program and cohort fixed effects. In Panel C, all regressions include the share of international students after residualizing it for program and cohort fixed effects. In Panel B, I weigh observations by the number of students from the program. The scale of question 4 in Panel A, and question 3 in Panels B and C is reversed compared to the original. Standard errors, which are clustered at the program level, are reported in parentheses. 
	\end{tablenotes}
\end{threeparttable} 
}
\end{table}

\pagebreak

\begingroup
\setlength{\tabcolsep}{10pt} 
\begin{table}[!htb]
\centering
{\footnotesize
\begin{threeparttable}
	\caption{Pre-emigration labor market outcomes of Dutch students}
    \label{emigration_missing} 
	
\begin{tabular}{llllll}
  \hline 
\hline \\[-2ex]  
 & Employed & Income & Entrepreneur & \% of foreign-born  \\ 
& & percentile & & co-workers  \\ 
  \hline
  \hline
Emigrated & -0.151 & -0.110 & -0.007 & 0.037  \\ 
 & (0.003) & (0.004) & (0.001) & (0.001) \\ 
         \hline
  Mean & 0.933 & 0.614 & 0.032 & 0.118 \\ 
  N & 2,496,650 & 2,496,650 & 2,496,650 & 1,767,597 \\ 
    \hline 
\hline 
\end{tabular}

    \begin{tablenotes}
			\footnotesize
			\item \textit{Notes}: All regressions include program and cohort fixed effects. This table reports coefficients from regressions of pre-emigration labor market outcomes on an indicator for whether the student subsequently emigrated 15 years after enrollment. Outcomes are measured in the year before the student's last observed year in the Netherlands. Mean refers to the sample mean of the dependent variable among non-emigrants. The sample contains outcomes for emigrants at the individual level measured once and for non-emigrants at the individual level measured annually. Standard errors, which are clustered at the program level, are reported in parentheses. 
	\end{tablenotes}
\end{threeparttable} 
}
\end{table}
\endgroup

\pagebreak

\begingroup
\setlength{\tabcolsep}{10pt} 
\begin{table}[!htb]
\centering
{\footnotesize
\begin{threeparttable}
	\caption{Bounding analyses for labor market outcomes of Dutch students}
    \label{emigration_imputation} 
	
\begin{tabular}{llllll}
  \hline 
\hline \\[-2ex]  
  \textbf{Panel A}: Lower bound & Employed & Income & Entrepreneur & \% of foreign-born  \\ 
& & percentile & & co-workers  \\ 
\hline
  \hline
International share& -0.003 & 0.000 & 0.000 & 0.000  \\ 
 in 10\%-points  & (0.001) & (0.002) & (0.001) & (0.001) \\ 
         \hline
  Mean & 0.882 & 0.688 & 0.067 & 0.114 \\ 
  N & 617,094 & 617,094 & 617,094 & 458,709 \\ 
    \hline
    \hline \\[-2ex] 
      \textbf{Panel B}: Upper bound & & & & \\ 
  \hline
  \hline
International share & -0.001 & 0.001 & 0.002 & 0.002  \\ 
in 10\%-points  & (0.001) & (0.001) & (0.001) & (0.001) \\ 
         \hline
  Mean & 0.947 & 0.733 & 0.133 & 0.132 \\ 
  N & 617,094 & 617,094 & 617,094 & 458,709 \\ 
    \hline
    \hline \\[-2ex] 
      \textbf{Panel C}: Pre-emigration  & & & & \\ 
  \hline
  \hline
International share  & -0.002 & -0.001 & 0.000 & 0.001  \\ 
in 10\%-points & (0.001) & (0.002) & (0.001) & (0.001) \\ 
         \hline
  Mean & 0.935 & 0.706 & 0.070 & 0.123 \\ 
  N & 606,230 & 606,230 & 606,230 & 441,006 \\ 
    \hline 
\hline 
\end{tabular}

    \begin{tablenotes}
			\footnotesize
			\item \textit{Notes}: All regressions include program and cohort fixed effects. This table reports estimates of the effect of exposure to international students on natives’ labor market outcomes, addressing missing outcomes for emigrants. Panel A presents lower bounds, imputing zeros for binary outcomes (employment and entrepreneurship) and values from the bottom 10\% of peers in emigrants’ programs for continuous outcomes (income percentile and the share of foreign-born co-workers). Panel B presents upper bounds, imputing ones for binary outcomes and values from the top 10\% of peers for continuous outcomes. Panel C presents estimates imputing emigrants’ outcomes from the year before their last observed year in the Netherlands. Outcomes are measured 15 years after enrollment. Mean refers to the sample mean of the dependent variable among non-emigrants. Standard errors, which are clustered at the program level, are reported in parentheses. 
	\end{tablenotes}
\end{threeparttable} 
}
\end{table}
\endgroup

\pagebreak

\begingroup
\setlength{\tabcolsep}{9pt} 
\begin{table}[!htb]
\centering
{\footnotesize
\begin{threeparttable}
	\caption{The effect of exposure to international students on the educational outcomes of native students}
    \label{results_education} 
	\begin{tabular}{llll}
  \hline
  \hline \\[-2ex] 
   &  Time to graduate & Bachelor's degree & Master's degree  \\ 
   & & completion & completion \\
\hline
  \hline
International share & -0.017 & 0.012 & 0.001 \\ 
in 10\%-points & (0.026) & (0.007) & (0.004) \\ 
         \hline
  Mean & 4.613 & 0.786 & 0.366 \\ 
  N & 782,590 & 1,037,347 & 1,037,347 \\
      \hline
   \hline
\end{tabular}

    \begin{tablenotes}
			\footnotesize
			\item \textit{Notes}: All regressions include program and cohort fixed effects. Standard errors, which are clustered at the program level, are reported in parentheses. 
	\end{tablenotes}
\end{threeparttable} 
}
\end{table}
\endgroup

\pagebreak

\begingroup
\setlength{\tabcolsep}{9pt} 
\begin{table}[!htb]
\centering
{\footnotesize
\begin{threeparttable}
	\caption{The effect of exposure to international students on ``native flight''}
    \label{native_flight_stem} 
	\begin{tabular}{llll}
  \hline
  \hline \\[-2ex] 
& Dropped out & Switched program & Switched university   \\ 
 \textbf{Panel A} & during 1st year & at the same university & to the same program  \\ 
\hline
  \hline
International share & -0.003 & 0.002 & 0.000  \\ 
in 10\%-points & (0.006) & (0.001) & (0.000) \\ 
         \hline
  Mean & 0.264 & 0.045 & 0.016 \\ 
  N & 1,037,347 & 1,037,347 & 1,037,347  \\
  \hline
    \hline \\[-2ex] 
 & Dropped out & Switched program & Switched university   \\ 
\textbf{Panel B}  & during 1st year & at the same university & to the same program  \\ 
\hline
  \hline
International share & -0.007 & 0.002 & 0.000  \\ 
in 10\%-points & (0.007) & (0.001) & (0.000) \\ 
x $\mathds{1}$ [STEM = 1] & 0.026 & 0.001 & 0.000 \\
& (0.009) & (0.003) & (0.002) \\
         \hline
  N & 1,037,347 & 1,037,347 & 1,037,347 \\
      \hline
   \hline
\end{tabular}
    \begin{tablenotes}
			\footnotesize
			\item \textit{Notes}: All regressions include program and cohort fixed effects. Switched program at the same university refers to whether a student changed program but remained at the same university during the first year. Switched university to the same program refers to whether a student transferred to a different university but stayed in the same program during the first year. Standard errors, which are clustered at the program level, are reported in parentheses. 
	\end{tablenotes}
\end{threeparttable} 
}
\end{table}
\endgroup

\pagebreak

\setlength{\tabcolsep}{3pt} 
\begin{table}[!htb]
\hspace*{-1cm} 
\centering
{\scriptsize
\begin{threeparttable}
	\caption{Heterogeneous effects of exposure to international students on the outcomes of native students 15 years post-enrollment}
    \label{results_hetero} 
	\begin{tabular}{lllllllll}
  \hline
  \hline \\[-2ex] 
\textbf{Panel A}: Sex & Cohabited with  & Married to & Emigrated & Employed & Income & Entrepreneur & \% of foreign-born  \\
& a non-native & a non-native & & & percentile & & co-workers  \\
  \hline
  \hline
International share & 0.006 & 0.002 & 0.002 & -0.002 & 0.001 & 0.000 & 0.000  \\ 
in 10\%-points & (0.001) & (0.001) & (0.001) & (0.001) & (0.002) & (0.001) & (0.001)  \\ 
x $\mathds{1}$ [Female = 1] & -0.001 & -0.001 & 0.001 & 0.001 & 0.000 & 0.000 & 0.001  \\ 
& (0.002) & (0.001) & (0.002) & (0.001) & (0.002) & (0.002) & (0.001)  \\ 
         \hline
  Mean for males & 0.115 & 0.044 & 0.072 & 0.953 & 0.698 & 0.070 & 0.130 \\ 
  Mean for females & 0.089 & 0.026 & 0.058 & 0.933 & 0.737 & 0.075 & 0.114  \\ 
  N & 605,367 & 587,676 & 617,094 & 576,961 & 576,961 & 576,961 & 427,687  \\ 
  \hline
  \hline \\[-2ex] 
\textbf{Panel B}: Field of study & & & & & & &  \\ 
  \hline
  \hline
International share & 0.007 & 0.002 & 0.003 & -0.001 & 0.000 & 0.000 & 0.001  \\ 
in 10\%-points & (0.001) & (0.001) & (0.001) & (0.001) & (0.002) & (0.001) & (0.001)  \\ 
x $\mathds{1}$ [STEM = 1] & -0.010 & -0.001 & -0.005 & 0.000 & 0.005 & 0.002 & -0.004  \\ 
& (0.004) & (0.003) & (0.004) & (0.002) & (0.004) & (0.003) & (0.002)  \\ 
         \hline
  Mean for non-STEM & 0.098 & 0.033 & 0.060 & 0.941 & 0.717 & 0.077 & 0.117  \\ 
  Mean for STEM & 0.115 & 0.043 & 0.084 & 0.951 & 0.717 & 0.055 & 0.139  \\ 
  N & 605,367 & 587,676 & 617,094 & 576,961 & 576,961 & 576,961 & 427,687  \\  
  \hline
  \hline \\[-2ex] 
\textbf{Panel C}: Share of females &  & & & &  &  & \\ 
  \hline
  \hline
International share & 0.006 & 0.002 & 0.003 & -0.001 & 0.001 & 0.000 & 0.001  \\ 
in 10\%-points & (0.001) & (0.001) & (0.001) & (0.001) & (0.002) & (0.001) & (0.001)  \\ 
x $\mathds{1}$ [Share is below 20\% = 1] & -0.011 & -0.002 & -0.005 & -0.006 & 0.001 & 0.000 & -0.002  \\ 
& (0.006) & (0.004) & (0.005) & (0.003) & (0.004) & (0.003) & (0.004)  \\ 
         \hline
  Mean for above 20\% & 0.100 & 0.033 & 0.063 & 0.942 & 0.714 & 0.077 & 0.119  \\ 
  Mean for below 20\% & 0.116 & 0.047 & 0.084 & 0.955 & 0.740 & 0.042 & 0.142  \\ 
  N & 605,367 & 587,676 & 617,094 & 576,961 & 576,961 & 576,961 & 427,687  \\ 
  \hline
  \hline \\[-2ex] 
\textbf{Panel D}: Program size & & & & & & & &  \\ 
  \hline
  \hline
International share & 0.008 & 0.001 & 0.004 & -0.001 & 0.000 & 0.000 & 0.002  \\ 
in 10\%-points & (0.001) & (0.001) & (0.001) & (0.001) & (0.002) & (0.001) & (0.001)  \\ 
x $\mathds{1}$ [Size is below average = 1] & -0.007  & 0.001 & -0.006 & 0.001 & -0.001 & 0.001 & -0.003 \\ 
& (0.003) &  (0.002) & (0.002) & (0.002) & (0.003) & (0.002) & (0.001)  \\ 
         \hline
  Mean for above average & 0.098 & 0.033 & 0.061 & 0.947 & 0.732 & 0.072 & 0.119 &  \\ 
  Mean for below average & 0.111 & 0.040 & 0.075 & 0.935 & 0.683 & 0.072 & 0.129  \\ 
  N & 605,367 & 587,676 & 617,094 & 576,961 & 576,961 & 576,961 & 427,687  \\  
    \hline
   \hline
\end{tabular}

    \begin{tablenotes}
			\scriptsize
			\item \textit{Notes}: All regressions include program and cohort fixed effects as well as their interactions with an indicator variable. A non-native is defined as a person without Dutch nationality. In Panel D, the indicator equals 1 if the program size is smaller than the average, which is 137. Standard errors, which are clustered at the program level, are reported in parentheses. 
	\end{tablenotes}
\end{threeparttable} 
}
\end{table}

\pagebreak

\begin{table}[!htbp]
\hspace*{-1.5cm} 
\centering
{\scriptsize
\begin{threeparttable}
	\caption{Sensitivity to alternative definitions of international students}
    \label{sensitivity_1} 
	\begin{tabular}{lllllllll}
  \hline
  \hline \\[-2ex] 
\textbf{Panel A}: Cohabited with a non-native & (1) & (2) & (3) & (4) & (5) & (6) & (7) & (8) \\
\hline
  \hline
International share & 0.006 & 0.006 & 0.005 & 0.006 & 0.005 & 0.005 & 0.008 & 0.002 \\
in 10\%-points & (0.001) & (0.001) & (0.001) & (0.001) & (0.001) & (0.001) & (0.002) & (0.006) \\
  \hline
  N & 605,367 & 605,367 & 605,367 & 605,357 & 605,367 & 605,367 & 605,367 & 605,367 \\
    \hline
  \hline \\[-2ex] 
\multicolumn{9}{l}{\textbf{Panel B}: Married to a non-native} \\
\hline
  \hline
International share & 0.001 & 0.002 & 0.001 & 0.000 & 0.001 & 0.002 & 0.001 & 0.003 \\
in 10\%-points & (0.001) & (0.001) & (0.001) & (0.001) & (0.001) & (0.001) & (0.001) & (0.003) \\
  \hline
  N & 587,676 & 587,676 & 587,676 & 587,666 & 587,676 & 587,676 & 587,676 & 587,676 \\
  \hline
  \hline \\[-2ex] 
  \multicolumn{9}{l}{\textbf{Panel C}: Emigrated}  \\
\hline
  \hline
International share  & 0.003 & 0.003 & 0.003 & 0.003 & 0.003 & 0.002 & 0.004 & -0.001 \\
in 10\%-points  & (0.001) & (0.001) & (0.001) & (0.001) & (0.001) & (0.001) & (0.001) & (0.005) \\
  \hline
  N  & 617,094 & 617,094 & 617,094 & 617,084 & 617,094 & 617,094 & 617,094 & 617,094 \\
  \hline
  \hline \\[-2ex] 
  \multicolumn{9}{l}{\textbf{Panel D}: Employed} \\
\hline
  \hline
International share & -0.001 & -0.001 & -0.001 & -0.001 & -0.001 & -0.002 & -0.002 & 0.000 \\ 
in 10\%-points & (0.001) & (0.001) & (0.001) & (0.001) & (0.001) & (0.001) & (0.001) & (0.004) \\
  \hline
  N & 576,961 & 576,961 & 576,961 & 576,951 & 576,961 & 576,961 & 576,961 & 576,961 \\
  \hline
  \hline \\[-2ex] 
\multicolumn{9}{l}{\textbf{Panel E}: Income percentile} \\
\hline
  \hline
International share & 0.001 & 0.000 & -0.001 & 0.000 & -0.001 & -0.003 & 0.000 & 0.005 \\
in 10\%-points & (0.001) & (0.001) & (0.001) & (0.002) & (0.001) & (0.001) & (0.002) & (0.006) \\
  \hline
  N & 576,961 & 576,961 & 576,961 & 576,951 & 576,961 & 576,961 & 576,961 & 576,961 \\
  \hline
  \hline \\[-2ex] 
\multicolumn{9}{l}{\textbf{Panel F}: Entrepreneur} \\
\hline
  \hline
International share & 0.000 & 0.000 & 0.001 & 0.001 & 0.001 & 0.002 & 0.000 & 0.003 \\
in 10\%-points & (0.001) & (0.001) & (0.001) & (0.001) & (0.001) & (0.001) & (0.001) & (0.004) \\
  \hline
  N & 576,961 & 576,961 & 576,961 & 576,951 & 576,961 & 576,961 & 576,961 & 576,961 \\
  \hline
  \hline \\[-2ex] 
\multicolumn{9}{l}{\textbf{Panel G}: \% of foreign-born co-workers} \\
\hline
  \hline
International share & 0.001 & 0.001 & 0.001 & 0.001 & 0.001 & 0.002 & 0.001 & -0.004 \\
in 10\%-points & (0.001) & (0.001) & (0.001) & (0.001) & (0.001) & (0.001) & (0.001) & (0.003) \\
  \hline
  N & 427,687 & 427,687 & 427,687 & 427,687 & 427,687 & 427,687 & 427,687 & 427,687 \\
         \hline 
        \hline
  \multicolumn{9}{l}{\textit{Definition of international students is based on}} \\
  ~~Nationality and high school completion & \checkmark & & & & & & & \\
  ~~Nationality & & \checkmark & & & & & & \\
  ~~High school completion & & & \checkmark & & & & & \\
  ~~Residency before enrollment & & & & \checkmark & & & & \\
  ~~Country of birth & & & & & \checkmark & & & \\
  ~~Parent's country of birth & & & & & & \checkmark & & \\
  ~~EEA+ nationality  & & & & & & & \checkmark & \\
  ~~Non-EEA+ nationality  & & & & & & & & \checkmark \\
      \hline
   \hline
\end{tabular}

    \begin{tablenotes}
			\scriptsize
			\item \textit{Notes}: All regressions include program and cohort fixed effects. A non-native is defined as a person without Dutch nationality. EEA+ nationality refers to individuals from the European Economic Area (the EU, Iceland, Liechtenstein, and Norway) plus Switzerland. Standard errors, which are clustered at the program level, are reported in parentheses.  
	\end{tablenotes}
\end{threeparttable} 
}
\end{table}

\pagebreak

\begin{table}[!htbp]
\hspace*{-1.5cm} 
\centering
{\scriptsize
\begin{threeparttable}
	\caption{Sensitivity to alternative definitions of international students}
    \label{sensitivity_2} 
	\begin{tabular}{lllllllll}
  \hline
  \hline \\[-2ex] 
\textbf{Panel A}: Satisfaction with the level of & (1) & (2) & (3) & (4) & (5) & (6) & (7) & (8)  \\
\multicolumn{9}{l}{internationalization} \\
\hline
  \hline
International share  & 0.053 & 0.056 & 0.052 & 0.054 & 0.051 & 0.046 & 0.061 & 0.101 \\
in 10\%-points  & (0.019) & (0.019) & (0.018) & (0.021) & (0.019) & (0.020) & (0.021) & (0.071) \\
  \hline
  N  & 123,038 & 123,038 & 123,038 & 123,038 & 123,038 & 123,038 & 123,038 & 123,038 \\
  \hline
  \hline \\[-2ex] 
\multicolumn{9}{l}{\textbf{Panel B}: Satisfaction with the level of encouragement to learn about other cultures} \\
\hline
  \hline
International share & 0.038 & 0.042 & 0.028 & 0.030 & 0.038 & 0.047 & 0.033 & 0.140 \\
in 10\%-points  & (0.019) & (0.019) & (0.019) & (0.021) & (0.019) & (0.020) & (0.022) & (0.065) \\
  \hline
N  & 123,038 & 123,038 & 123,038 & 123,038 & 123,038 & 123,038 & 123,038 & 123,038 \\
  \hline
  \hline \\[-2ex] 
\multicolumn{9}{l}{\textbf{Panel C}: Social security rights for foreigners} \\
\hline
  \hline
International share  & 0.238 & 0.230 & 0.204 & 0.222 & 0.235 & 0.218 & 0.202 & 0.768 \\
in 10\%-points  & (0.095) & (0.095) & (0.094) & (0.093) & (0.095) & (0.082) & (0.097) & (0.553) \\
  \hline
  N  & 1,180 & 1,180 & 1,180 & 1,180 & 1,180 & 1,180 & 1,180 & 1,180 \\
  \hline
  \hline \\[-2ex] 
\multicolumn{9}{l}{\textbf{Panel D}: Foreigners in neighborhoods} \\
\hline
  \hline
International share  & -0.138 & -0.137 & -0.125 & -0.128 & -0.140 & -0.137 & -0.121 & -0.592 \\
in 10\%-points  & (0.062) & (0.061) & (0.059) & (0.064) & (0.066) & (0.068) & (0.054) & (0.415) \\
  \hline
  N  & 1,180 & 1,180 & 1,180 & 1,180 & 1,180 & 1,180 & 1,180 & 1,180 \\
  \hline
  \hline \\[-2ex] 
\multicolumn{9}{l}{\textbf{Panel E}: European unification} \\
\hline
  \hline
International share  & 0.239 & 0.237 & 0.216 & 0.229 & 0.244 & 0.232 & 0.245 & 0.597 \\
in 10\%-points  & (0.068) & (0.066) & (0.067) & (0.066) & (0.066) & (0.073) & (0.059) & (0.366) \\
  \hline
  N  & 1,166 & 1,166 & 1,166 & 1,166 & 1,166 & 1,166 & 1,166 & 1,166 \\
         \hline 
        \hline
  \multicolumn{9}{l}{\textit{Definition of international students is based on}} \\
  ~~Nationality and high school completion & \checkmark & & & & & & & \\
  ~~Nationality & & \checkmark & & & & & & \\
  ~~High school completion & & & \checkmark & & & & & \\
  ~~Residency before enrollment & & & & \checkmark & & & & \\
  ~~Country of birth & & & & & \checkmark & & & \\
  ~~Parent's country of birth & & & & & & \checkmark & & \\
  ~~EEA+ nationality  & & & & & & & \checkmark & \\
  ~~Non-EEA+ nationality  & & & & & & & & \checkmark \\
      \hline
   \hline
\end{tabular}

    \begin{tablenotes}
			\scriptsize
			\item \textit{Notes}: All regressions include program and cohort fixed effects. EEA+ nationality refers to individuals from the European Economic Area (the EU, Iceland, Liechtenstein, and Norway) plus Switzerland. Standard errors, which are clustered at the program level, are reported in parentheses.  
	\end{tablenotes}
\end{threeparttable} 
}
\end{table}

\pagebreak

\begingroup
\setlength{\tabcolsep}{4pt} 
\begin{table}[!htb]
\hspace*{-2cm} 
\centering
{\footnotesize
\begin{threeparttable}
	\caption{The effect of exposure to international students on the outcomes of native students 10 and 25 years post-enrollment}
    \label{results_different_periods} 
	\begin{tabular}{lllllllll}
  \hline
  \hline \\[-2ex] 
\textbf{Panel A}: 10 years & Cohabited with  & Married to & Emigrated & Employed & Income & Entrepreneur & \% of foreign-born  \\
post-enrollment & a non-native & a non-native & & & percentile & & co-workers  \\ 
  \hline
  \hline
International share & 0.007 & 0.001 & 0.004 & 0.001 & 0.003 & -0.001 & 0.001  \\ 
in 10\%-points & (0.002) & (0.000) & (0.001) & (0.001) & (0.002) & (0.001) & (0.001)  \\ 
         \hline
 Mean & 0.077 & 0.015 & 0.041 & 0.948 & 0.679 & 0.039 & 0.129  \\ 
 Cohorts & 1988-2013 & 1988-2013 & 1988-2013 & 1993-2013 & 1993-2013 & 1993-2013 & 1996-2013  \\ 
  N & 787,329 & 773,073 & 801,346 & 614,337 & 614,337 & 614,337 & 491,117  \\ 
  \hline 
\hline \\[-2ex] 
 \textbf{Panel B}: 25 years  & & & & & & & \\
post-enrollment & & & & & & & \\
  \hline
  \hline
International share & 0.004 & 0.003 & 0.000 & 0.001 & 0.000 & -0.001 & 0.000  \\ 
in 10\%-points & (0.002) & (0.001) & (0.002) & (0.001) & (0.002) & (0.002) & (0.001)  \\ 
         \hline
  Mean & 0.118 & 0.053 & 0.082 & 0.932 & 0.720 & 0.129 & 0.121  \\ 
  Cohorts & 1988-1998 & 1988-1998 & 1988-1998 & 1988-1998 & 1988-1998 & 1988-1998 & 1988-1998  \\ 
  N & 311,271 & 300,021 & 315,779 & 289,898 & 289,898 & 289,898 & 236,121 \\ 
    \hline
   \hline
\end{tabular}

    \begin{tablenotes}
			\footnotesize
			\item \textit{Notes}: All regressions include program and cohort fixed effects. A non-native is defined as a person without Dutch nationality. Standard errors, which are clustered at the program level, are reported in parentheses. 
	\end{tablenotes}
\end{threeparttable} 
}
\end{table}

\pagebreak

\begin{table}[!htbp]
\hspace*{-2.5cm} 
\centering
{\scriptsize
\begin{threeparttable}
	\caption{Robustness to alternative models and samples}
    \label{robustness_1} 
	\begin{tabular}{lllllllllll}
  \hline
  \hline \\[-2ex] 
\textbf{Panel A}: Cohabited with a non-native & (1) & (2) & (3) & (4) & (5) & (6) & (7) & (8) & (9) & (10) \\
\hline
  \hline
International share & 0.006 & 0.006 & 0.006 & 0.006 & 0.006 & 0.006 & 0.006 & 0.008 & 0.006 & 0.007 \\ 
in 10\%-points & (0.001) & (0.001) & (0.001) & (0.001) & (0.001) & (0.001) & (0.001) & (0.001) & (0.001) & (0.001) \\
  \hline
  N & 605,367 & 605,367 & 605,367 & 605,367 & 605,367 & 605,367 & 605,367 & 637,725 & 615,410 & 650,182  \\
    \hline
  \hline \\[-2ex] 
\multicolumn{11}{l}{\textbf{Panel B}: Married to a non-native} \\
\hline
  \hline
International share & 0.001 & 0.001 & 0.001 & 0.002 & 0.001 & 0.002 & 0.002 & 0.002 & 0.001 & 0.002 \\ 
in 10\%-points & (0.001) & (0.001) & (0.001) & (0.001) & (0.001) & (0.001) & (0.001) & (0.001) & (0.001) & (0.001)  \\
  \hline
  N & 587,676 & 587,676 & 587,676 & 587,676 & 587,676 & 587,676 & 587,676 & 617,035 & 597,498 & 629,126 \\
  \hline
  \hline \\[-2ex] 
  \multicolumn{11}{l}{\textbf{Panel C}: Emigrated} \\ 
\hline
  \hline
International share & 0.003 & 0.003 & 0.003 & 0.003 & 0.003 & 0.002 & 0.002 & 0.005 & 0.002 & 0.004 \\
in 10\%-points & (0.001) & (0.001) & (0.001) & (0.001) & (0.001) & (0.001) & (0.001) & (0.001) & (0.001) & (0.001) \\
  \hline
  N  & 617,094 & 617,094 & 617,094 & 617,094 & 617,094 & 617,094 & 617,094 & 651,650 & 627,143 & 664,272 \\
    \hline
  \hline \\[-2ex] 
\multicolumn{11}{l}{\textbf{Panel D}: Employed} \\
\hline
  \hline
International share & -0.001 & -0.001 & -0.001 & -0.001 & -0.001 & -0.001 & -0.001 & -0.001 & -0.001 & -0.001  \\ 
in 10\%-points & (0.001) & (0.001) & (0.001) & (0.001) & (0.001) & (0.001) & (0.001) & (0.001) & (0.001) & (0.001)  \\
  \hline
  N & 576,961 & 576,961 & 576,961 & 576,961 & 576,961 & 576,961 & 576,961 & 605,238 & 586,693 & 617,177 \\
  \hline
  \hline \\[-2ex] 
\multicolumn{11}{l}{\textbf{Panel E}: Income percentile}  \\
\hline
  \hline
International share & 0.001 & 0.001 & 0.001 & 0.000 & 0.002 & 0.001 & 0.001 & 0.000 & 0.001 & 0.000  \\ 
in 10\%-points & (0.001) & (0.001) & (0.001) & (0.002) & (0.001) & (0.001) & (0.001) & (0.001) & (0.001) & (0.001)  \\
  \hline
  N & 576,961 & 576,961 & 576,961 & 576,961 & 576,961 & 576,961 & 576,961 & 605,238 & 586,693 & 617,177 \\
  \hline
  \hline \\[-2ex] 
\multicolumn{11}{l}{\textbf{Panel F}: Entrepreneur} \\
\hline
  \hline
International share & 0.000 & 0.000 & 0.000 & 0.000 & 0.000 & 0.000 & 0.000 & 0.001 & 0.000 & 0.001 \\ 
in 10\%-points & (0.001) & (0.001) & (0.001) & (0.001) & (0.001) & (0.001) & (0.001) & (0.001) & (0.001) & (0.001)  \\
  \hline
  N & 576,961 & 576,961 & 576,961 & 576,961 & 576,961 & 576,961 & 576,961 & 605,238 & 586,693 & 617,177 \\
    \hline
  \hline \\[-2ex] 
\multicolumn{11}{l}{\textbf{Panel G}: \% of foreign-born co-workers}  \\
\hline
  \hline
International share & 0.001 & 0.001 & 0.001 & 0.001 & 0.001 & 0.001 & 0.001 & 0.001 & 0.001 & 0.001 \\ 
in 10\%-points & (0.001) & (0.001) & (0.001) & (0.001) & (0.001) & (0.001) & (0.001) & (0.001) & (0.001) & (0.001)  \\
  \hline
  N & 427,687 & 427,687 & 427,687 & 427,687 & 427,687 & 427,687 & 427,687 & 447,157 & 432,395 & 453,029  \\
         \hline 
        \hline
  \textit{Controls} & & & & & & & & & & \\
  ~~Field specific linear time trend & & \checkmark & & & & & \checkmark & & & \\
  ~~Field specific quadratic time trend & & & \checkmark & & & & \checkmark & & & \\
  ~~Program size & & & & \checkmark & & & \checkmark & & & \\
  ~~Individual characteristics & & & & & \checkmark & & \checkmark & & & \\
  ~~Peer characteristics & & & & & & \checkmark & \checkmark & & & \\
  \textit{Sample} & & & & & & & & & & \\
  ~~First-generation immigrants & & & & & & & & \checkmark & & \checkmark  \\
  ~~Students older than 30 at enrollment & & & & & & & &  & \checkmark & \checkmark \\
      \hline
   \hline
\end{tabular}

    \begin{tablenotes}
			\scriptsize
			\item \textit{Notes}: All regressions include program and cohort fixed effects. A non-native is defined as a person without Dutch nationality. Individual characteristics include sex, age at enrollment, migration status, family size, pre-vocational diploma, gap year indicator. Peer characteristics include the share of females, average age at enrollment, the share of second generation migrants, average family size, the share with a pre-vocational diploma, the share with a gap year. Standard errors, which are clustered at the program level, are reported in parentheses.  
	\end{tablenotes}
\end{threeparttable} 
}
\end{table}

\pagebreak

\begin{table}[!htbp]
\hspace*{-2.5cm} 
\centering
{\scriptsize
\begin{threeparttable}
	\caption{Robustness to alternative models and samples}
    \label{robustness_2} 
	\begin{tabular}{lllllllllll}
  \hline
  \hline \\[-2ex] 
\textbf{Panel A}: Satisfaction with the level of & (1) & (2) & (3) & (4) & (5) & (6) & (7) & (8) & (9) & (10) \\ 
\multicolumn{11}{l}{internationalization} \\
\hline
  \hline
International share  & 0.053 & 0.053 & 0.053 & 0.053 & 0.053 & 0.052 & 0.052 & 0.053 & 0.054 & 0.053 \\
in 10\%-points & (0.019) & (0.019) & (0.019) & (0.020) & (0.019) & (0.019) & (0.019) & (0.019) & (0.019) & (0.019)  \\
  \hline
  N  & 123,038 & 123,038 & 123,038 & 123,038 & 123,038 & 123,038 & 123,038 & 129,064 & 123,918 & 130,256 \\
  \hline
  \hline \\[-2ex] 
\multicolumn{11}{l}{\textbf{Panel B}: Satisfaction with the level of encouragement to learn about other cultures} \\
\hline
  \hline
International share  & 0.038 & 0.038 & 0.038 & 0.037 & 0.038 & 0.043 & 0.042 & 0.038 & 0.037 & 0.037 \\
in 10\%-points & (0.019) & (0.019) & (0.019) & (0.019) & (0.019) & (0.020) & (0.020) & (0.019) & (0.019) & (0.019)  \\
  \hline
  N  & 123,038 & 123,038 & 123,038 & 123,038 & 123,038 & 123,038 & 123,038 & 129,064 & 123,918 & 130,256 \\
  \hline
  \hline \\[-2ex] 
\multicolumn{11}{l}{\textbf{Panel C}: Social security rights for foreigners} \\
\hline
  \hline
International share  & 0.238 & 0.238 & 0.238 & 0.256 & 0.226 & 0.260 & 0.262 & 0.230 & 0.238 & 0.230 \\
in 10\%-points & (0.095) & (0.095) & (0.097) & (0.106) & (0.093) & (0.081) & (0.098) & (0.091) & (0.095) & (0.091)  \\
  \hline
N  & 1,180 & 1,180 & 1,180 & 1,180 & 1,180 & 1,180 & 1,180 & 1,251 & 1,193 & 1,267 \\
  \hline
  \hline \\[-2ex] 
\multicolumn{11}{l}{\textbf{Panel D}: Foreigners in neighborhoods} \\
\hline
  \hline
International share  & -0.138 & -0.138 & -0.138 & -0.153 & -0.112 & -0.126 & -0.123 & -0.118 & -0.138 & -0.117 \\
in 10\%-points & (0.062) & (0.062) & (0.063) & (0.066) & (0.060) & (0.063) & (0.069) & (0.062) & (0.062) & (0.062)  \\
  \hline
N  & 1,180 & 1,180 & 1,180 & 1,180 & 1,180 & 1,180 & 1,180 & 1,251 & 1,193 & 1,267 \\
  \hline
  \hline \\[-2ex] 
\multicolumn{11}{l}{\textbf{Panel E}: European unification} \\
\hline
  \hline
International share  & 0.239 & 0.239 & 0.239 & 0.234 & 0.248 & 0.197 & 0.185 & 0.206 & 0.239 & 0.206 \\
in 10\%-points & (0.068) & (0.068) & (0.069) & (0.065) & (0.069) & (0.079) & (0.085) & (0.068) & (0.068) & (0.068)  \\
  \hline
  N  & 1,166 & 1,166 & 1,166 & 1,166 & 1,166 & 1,166 & 1,166 & 1,235 & 1,179 & 1,250 \\
         \hline 
        \hline
  \textit{Controls} & & & & & & & & & & \\
  ~~Field specific linear time trend & & \checkmark & & & & & \checkmark & & & \\
  ~~Field specific quadratic time trend & & & \checkmark & & & & \checkmark & & &  \\
  ~~Program size & & & & \checkmark & & & \checkmark & & & \\
  ~~Individual characteristics & & & & & \checkmark & & \checkmark & & & \\
  ~~Peer characteristics & & & & & & \checkmark & \checkmark & & &  \\
  \textit{Sample} & & & & & & & & & & \\
  ~~First-generation immigrants & & & & & & & & \checkmark & & \checkmark  \\
  ~~Students older than 30 at enrollment & & & & & & & &  & \checkmark & \checkmark \\
      \hline
   \hline
\end{tabular}

    \begin{tablenotes}
			\scriptsize
			\item \textit{Notes}: All regressions include program and cohort fixed effects. Individual characteristics include sex, age at enrollment, migration status, family size, pre-vocational diploma, gap year indicator. Peer characteristics include the share of females, average age at enrollment, the share of second generation migrants, average family size, the share with a pre-vocational diploma, the share with a gap year. Standard errors, which are clustered at the program level, are reported in parentheses.  
	\end{tablenotes}
\end{threeparttable} 
}
\end{table}

\pagebreak

\begin{table}[!htbp]
\hspace*{-3cm} 
\centering
{\scriptsize
\begin{threeparttable}
    \caption{Overview of estimates on the impact of international students in higher education}
        \label{literature}
        \hspace*{-2cm} 
        \begin{tabular}{p{2cm} p{1cm} p{4cm} p{1cm} l p{6cm} p{2cm}}
  \hline
  \hline \\[-2ex] 
Paper & Country & Data & Sample & Measure & Relevant outcomes & Results \\ 
\hline
  \hline \\[-2ex] 
\cite{Chevalier2020} & UK  & Administrative data from one university and a survey of graduates & \parbox[t]{7cm}{4,032 \\ 1,581} & 9\%-points (SD) & \parbox[t]{7.5cm}{Failed a course \\ Abroad 6 months post-graduation} & \parbox[t]{7cm}{0.029 (0.058) \\ 0.015 (0.032)} \\ 
 \hline \\[-2ex] 
\cite{Costas2023} & UK & Administrative data and a survey of graduates & \parbox[t]{7cm}{509,870 \\ 315,215 \\ 124,305} & 18.5\%-points (SD) & \parbox[t]{6.5cm}{Graduated from a university \\ Employed 6 months post-graduation \\ Log yearly earnings 6 months post-graduation} & \parbox[t]{7cm}{0.001 (0.006) \\ 0.001 (0.009) \\ 0.033 (0.015)}\\ 
  \hline \\[-2ex] 
\cite{Anelli2023} & US & Administrative data from one university & 16,828 & 4.4\%-points (SD) & \parbox[t]{6.5cm}{Time to final major declaration  \\ Expected earnings 11-15 years post-graduation} &  \parbox[t]{6.5cm}{0.100 (0.120) \\ 0.031 (0.086)}  \\ 
  \hline \\[-2ex] 
\cite{Rakesh2023} & US & Administrative data from one university & 29,246 & 10\%-points & Graduated from a university & -0.011 (0.006) \\ 
      \hline
   \hline
\end{tabular}
            \begin{tablenotes}
			 \footnotesize
			 \item \textit{Notes}: Measure denotes the unit used to quantify the share of international students. ``SD'' signifies that a paper quantified this measure in terms of standard deviations, while the term ``\%-points'' indicates the magnitude of these standard deviations. Relevant outcomes are those findings from the literature that can be directly compared to the outcomes of this study. 
	   \end{tablenotes}
    \end{threeparttable} 
}
\end{table}

\pagebreak

\section*{Online Appendix C: Figures}

\begin{figure}[!htb]
	\caption{Histogram of the share of international students by program $\times$ year of enrollment}
 \includegraphics[width=0.8\textwidth]{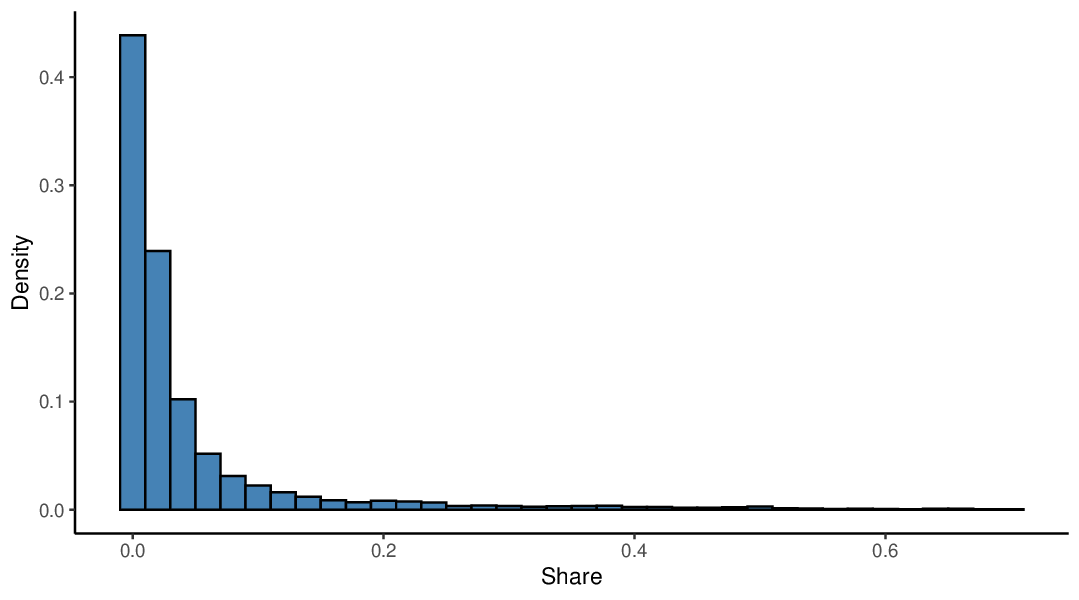} %
	\label{share} \\
	{\footnotesize \textit{Notes:} This histogram shows the plot of the share of international students calculated by program $\times$ year of enrollment. For display clarity, the top 1\% of the data is winsorized.}
\end{figure}

\begin{figure}[!htb]
	\caption{Histogram of the residualized share of international students by program $\times$ year of enrollment}
	\includegraphics[width=0.8\textwidth]{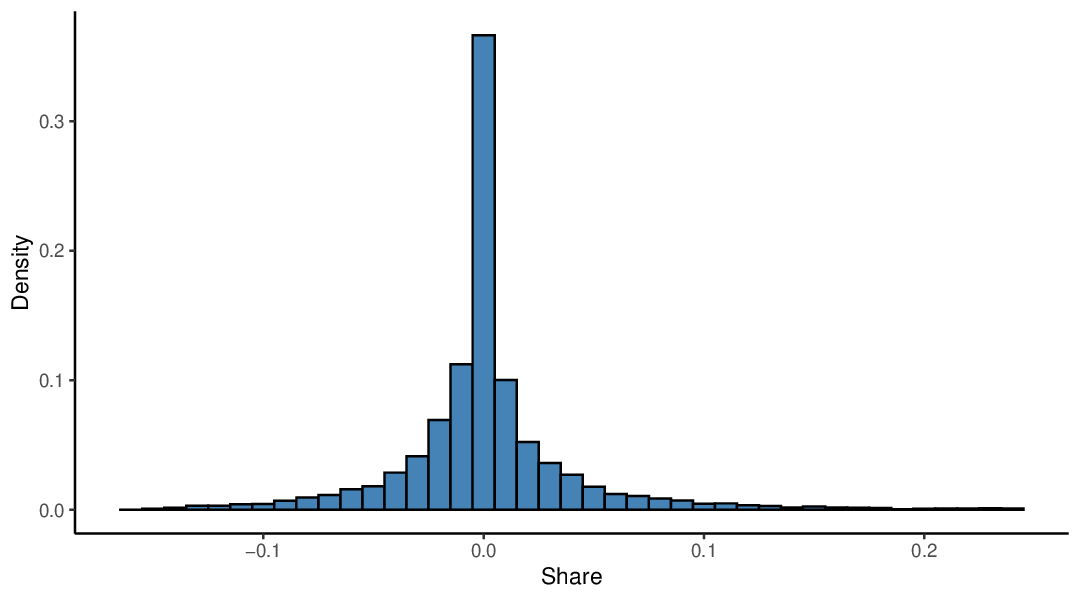} 
	\label{share_residualised} \\
	{\footnotesize \textit{Notes:} This histogram shows the plot of residuals of the share of international students calculated by program $\times$ year of enrollment after partialing out for program and cohort fixed effects. For display clarity, the top and bottom 1\% of the data are winsorized.}
\end{figure}

\pagebreak

\begin{figure}[!htb]
	\caption{Histogram of the share of international students by program $\times$ year of enrollment using the NSS sample}
 \includegraphics[width=0.8\textwidth]{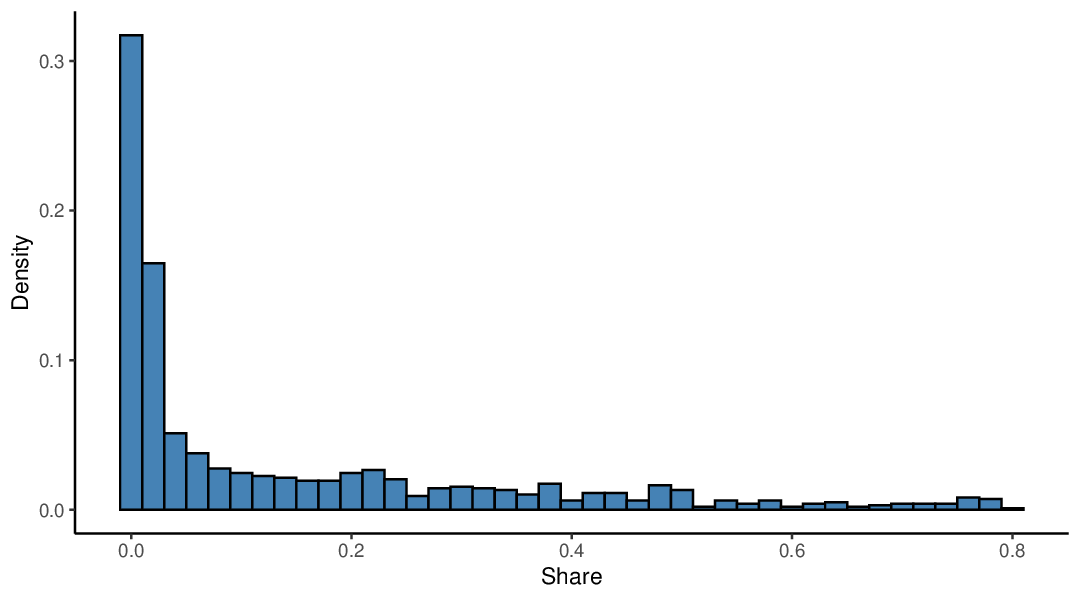}
	\label{share_nss} \\
	{\footnotesize \textit{Notes:} This histogram shows the plot of the share of international students calculated by program $\times$ year of enrollment. For display clarity, the top 1\% of the data is winsorized.}
\end{figure}

\begin{figure}[!htb]
	\caption{Histogram of the residualized share of international students by program $\times$ year of enrollment using the NSS sample}
	\includegraphics[width=0.8\textwidth]{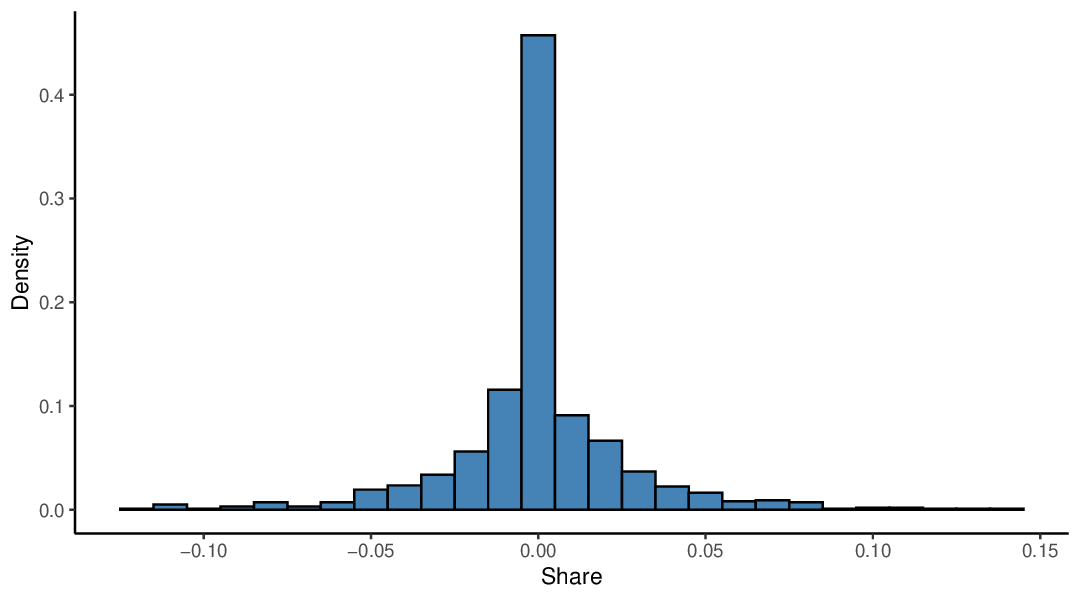} 
	\label{share_residualised_nss} \\
	{\footnotesize \textit{Notes:} This histogram shows the plot of residuals of the share of international students calculated by program $\times$ year of enrollment after partialing out for program and cohort fixed effects. For display clarity, the top and bottom 1\% of the data are winsorized.}
\end{figure}

\pagebreak

\begin{figure}[!htb]
\caption{Scatter plots}
\label{visual_1}
    \begin{subfigure}[b]{0.49\textwidth}
        \centering
        \includegraphics[width=\textwidth]{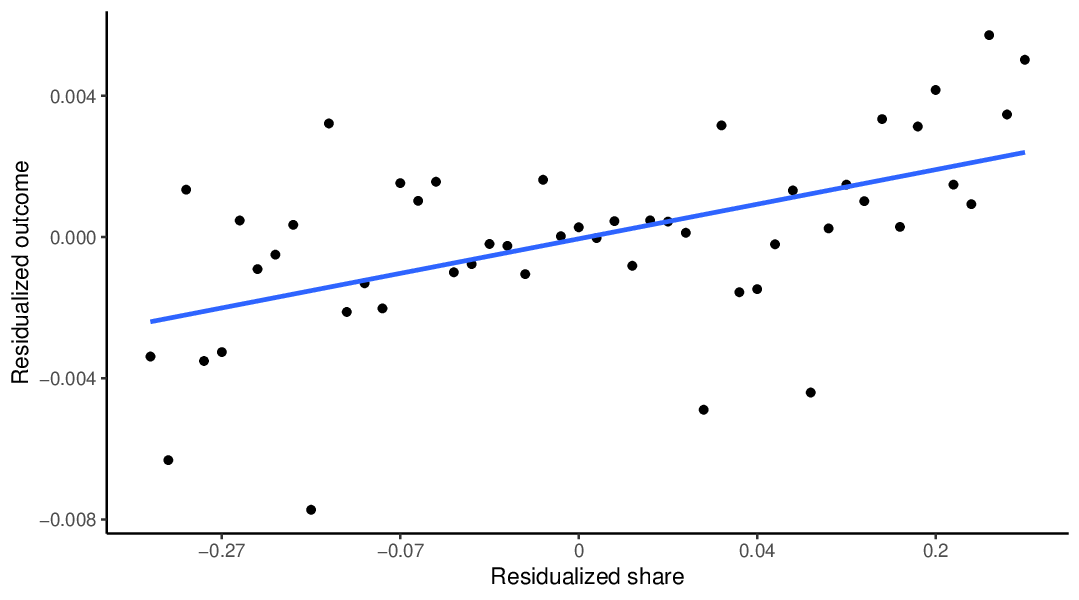}
        \caption{Cohabitation with a non-native}
    \end{subfigure}
    \hfill
    \begin{subfigure}[b]{0.49\textwidth}
        \centering
        \includegraphics[width=\textwidth]{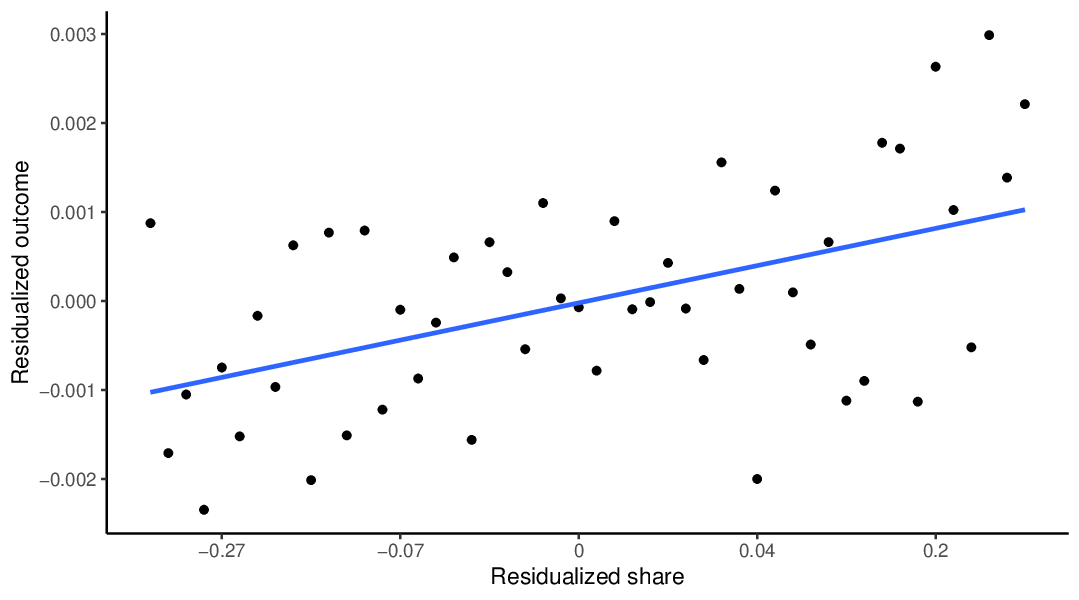}
        \caption{Marriage to a non-native}
    \end{subfigure}
    
    \vspace{1em}
    
    \begin{subfigure}[b]{0.49\textwidth}
        \centering
        \includegraphics[width=\textwidth]{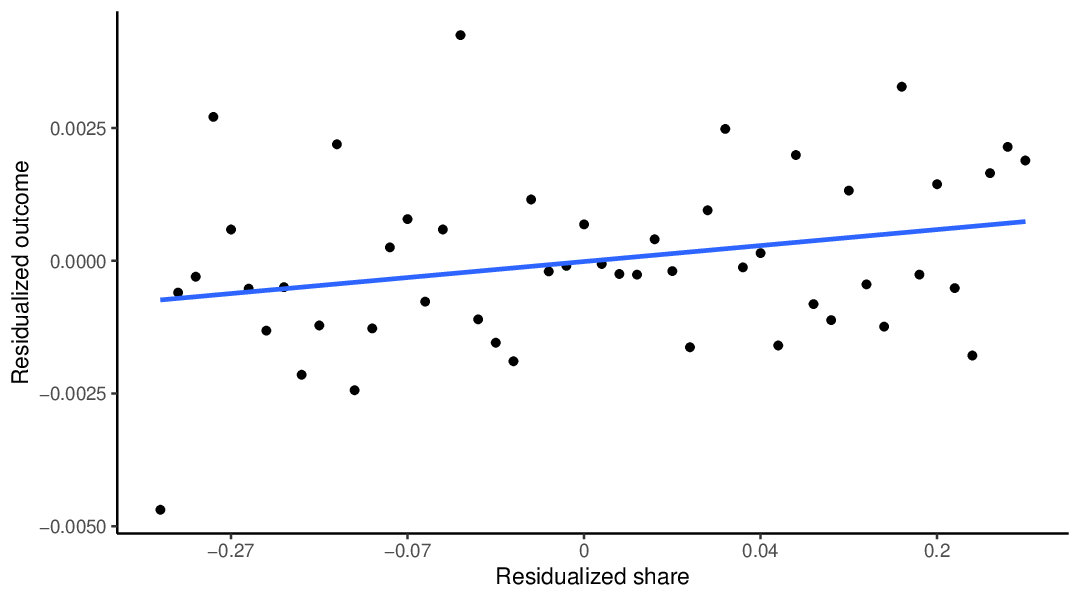}
        \caption{Emigration}
    \end{subfigure}
    \hfill
    \begin{subfigure}[b]{0.49\textwidth}
        \centering
        \includegraphics[width=\textwidth]{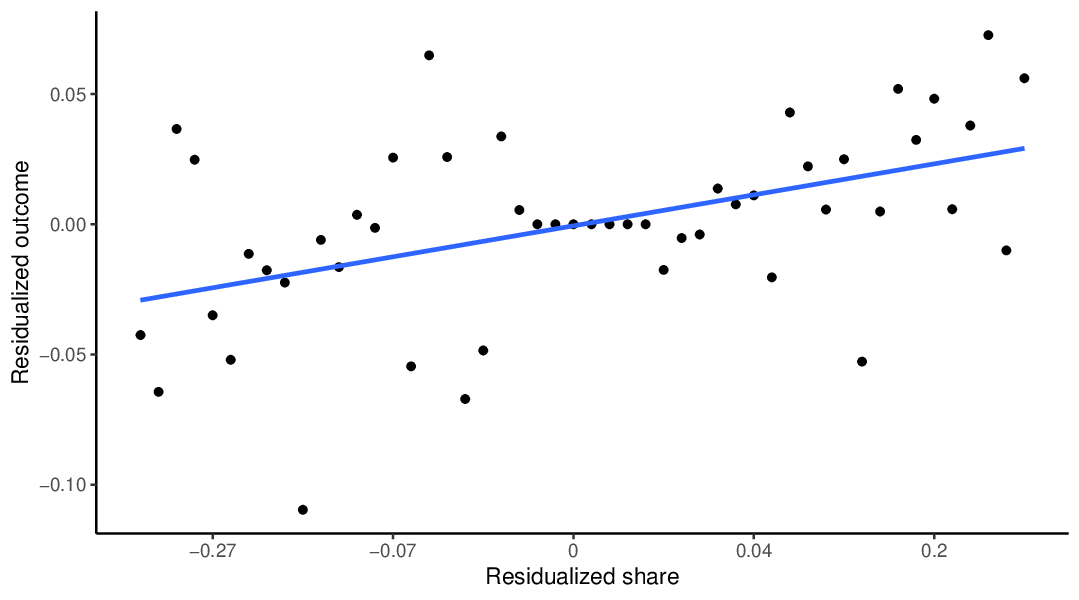}
        \caption{Satisfaction with internationalization}
    \end{subfigure}
    
    \textit{Notes:} The plots show the residualized outcome against the residualized share of international students after partialling out program and cohort fixed effects. The residualized share is divided into 50 equally sized bins. For each bin the average residualized outcome is shown against the average residualized share.
\end{figure}
\pagebreak

\begin{figure}[!htb]
\caption{Scatter plots}
\label{visual_2}
    \begin{subfigure}[b]{0.49\textwidth}
        \centering
        \includegraphics[width=\textwidth]{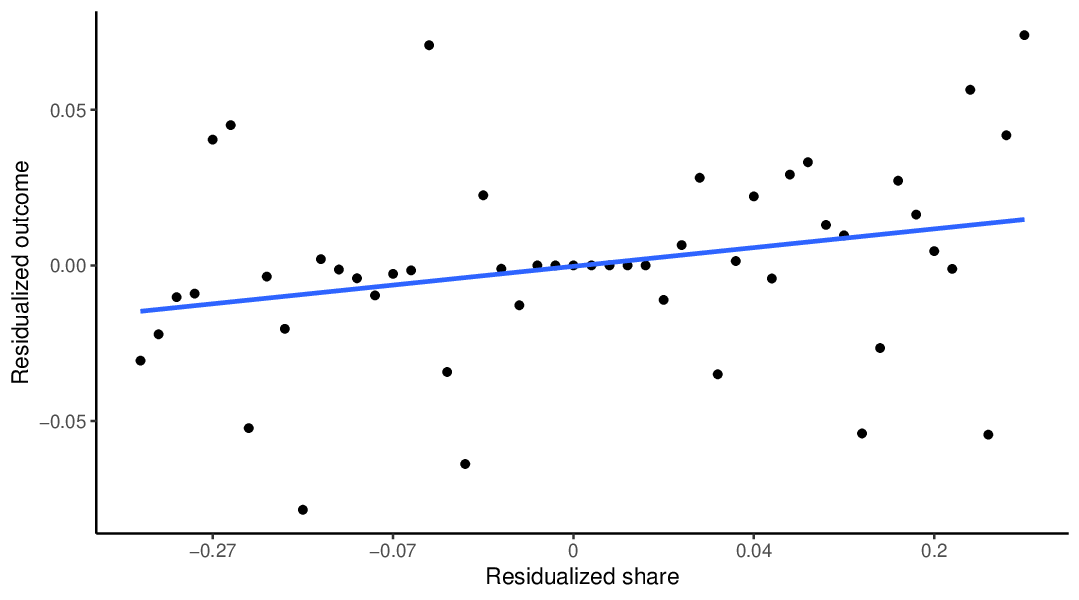}
        \caption{Satisfaction with encouragement to learn about other cultures}
    \end{subfigure}
    \hfill
    \begin{subfigure}[b]{0.49\textwidth}
        \centering
        \includegraphics[width=\textwidth]{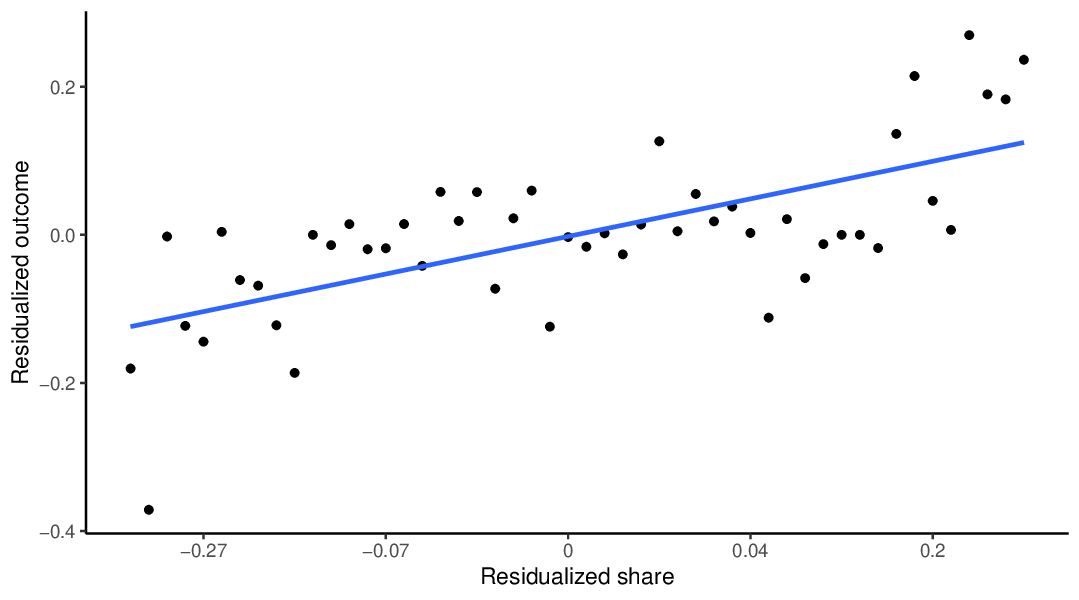}
        \caption{Same social security rights for foreigners}
    \end{subfigure}
    
    \vspace{1em}
    
    \begin{subfigure}[b]{0.49\textwidth}
        \centering
        \includegraphics[width=\textwidth]{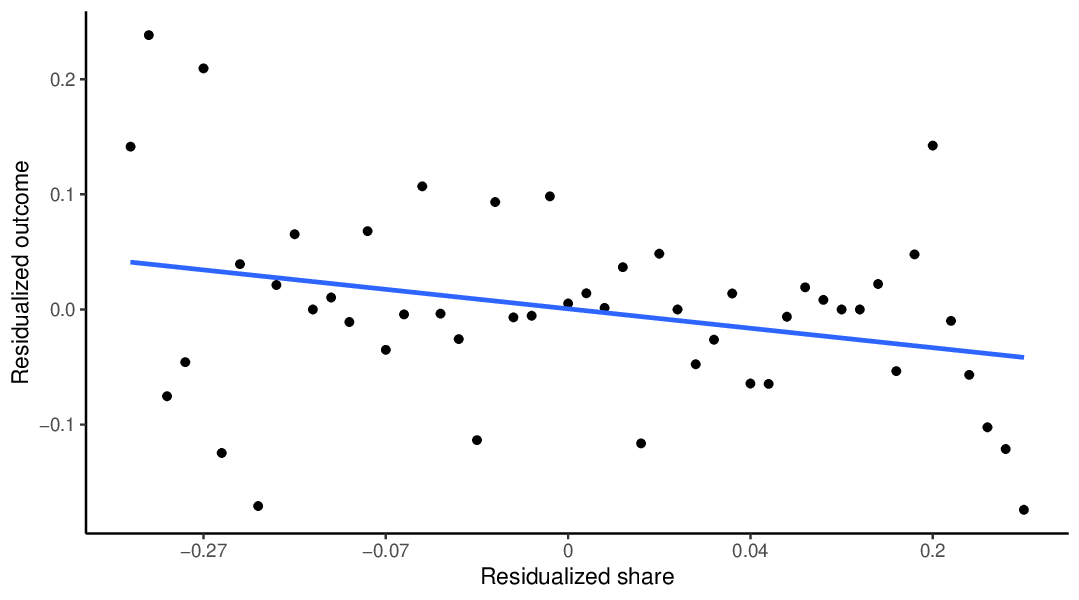}
        \caption{Foreigners in neighborhoods}
    \end{subfigure}
    \hfill
    \begin{subfigure}[b]{0.49\textwidth}
        \centering
        \includegraphics[width=\textwidth]{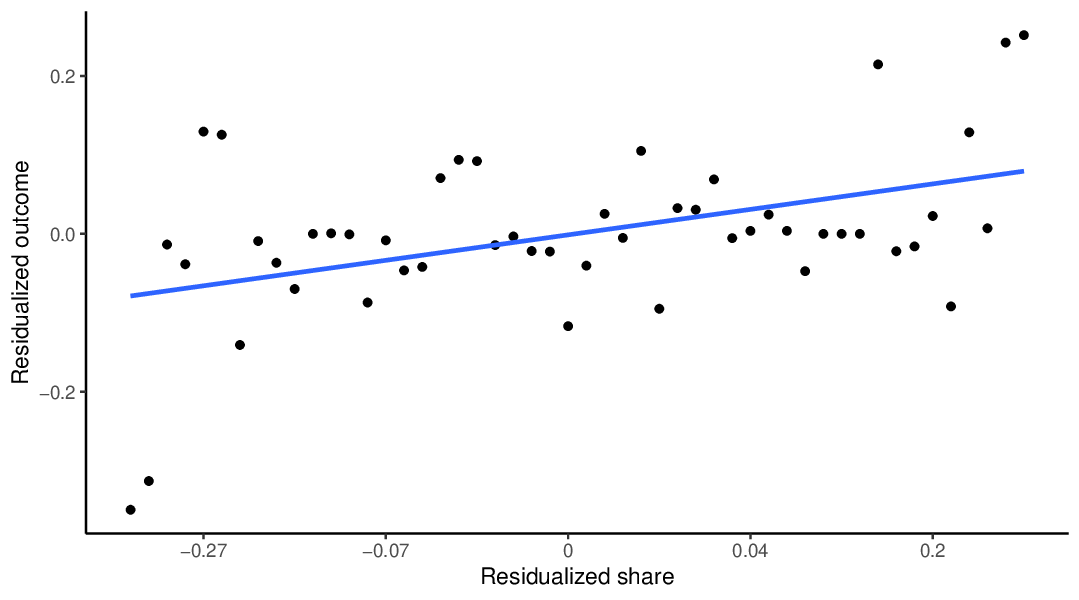}
        \caption{European unification}
    \end{subfigure}
    
    \textit{Notes:} The plots show the residualized outcome against the residualized share of international students after partialling out program and cohort fixed effects. The residualized share is divided into 50 equally sized bins. For each bin the average residualized outcome is shown against the average residualized share.
\end{figure}
\pagebreak

\begin{figure}[!htb]
\caption{Marginal effects of the share of international students}
\label{marginal_1}
	\begin{subfigure}[b]{0.49\textwidth}
		\includegraphics[width=\textwidth]{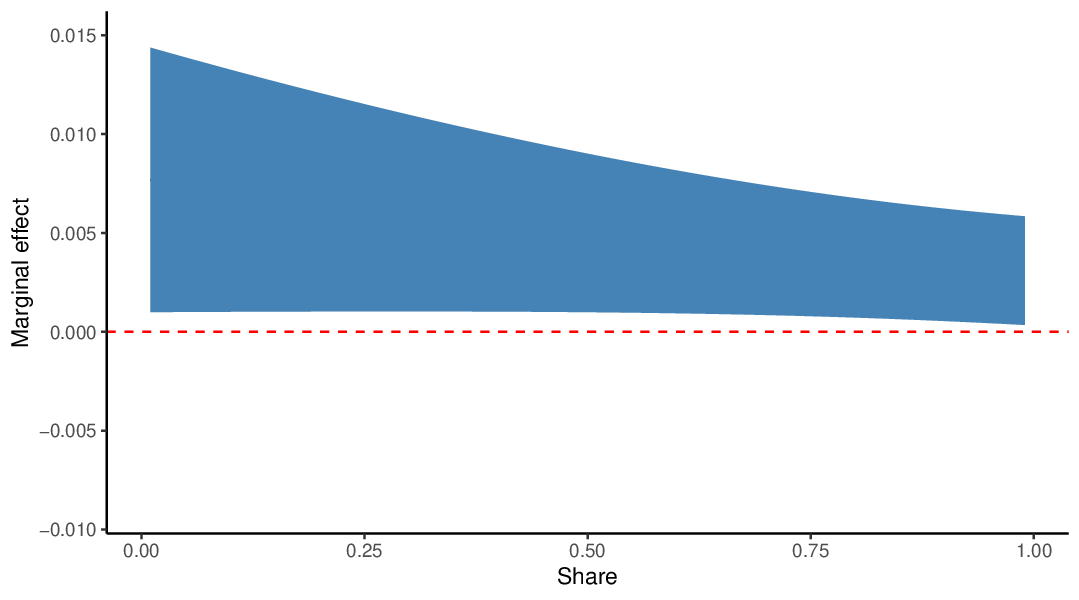}
		\caption{Cohabitation with a non-native}
		\label{marginal_cohabited}
	\end{subfigure}
	\hfill
	\begin{subfigure}[b]{0.49\textwidth}
		\includegraphics[width=\textwidth]{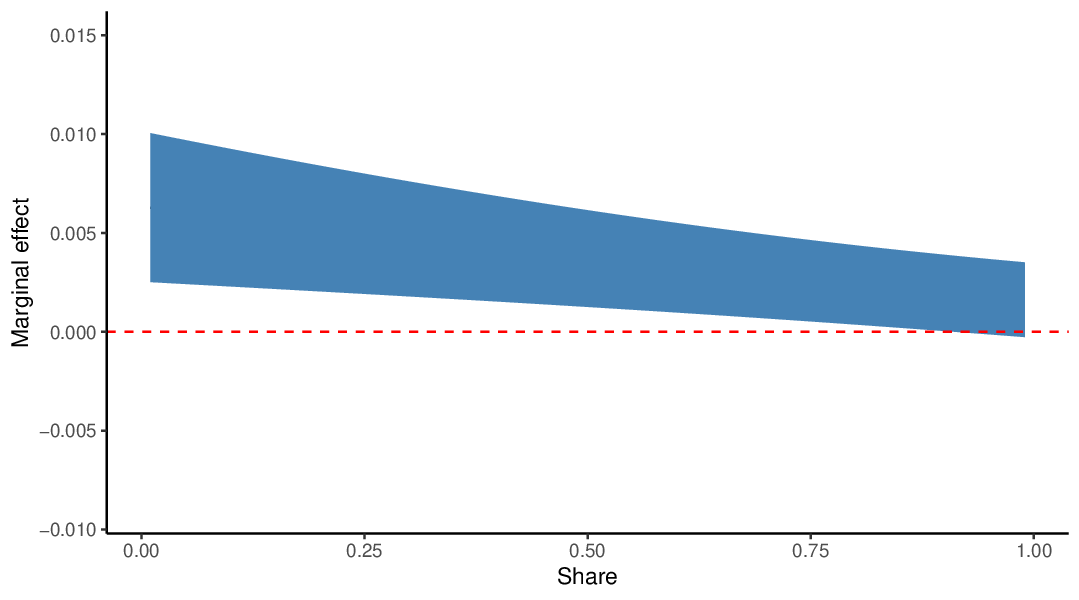}
		\caption{Marriage to a non-native}
		\label{marginal_married}
	\end{subfigure}
	
	\vspace{1em}
	
	\begin{subfigure}[b]{0.49\textwidth}
		\includegraphics[width=\textwidth]{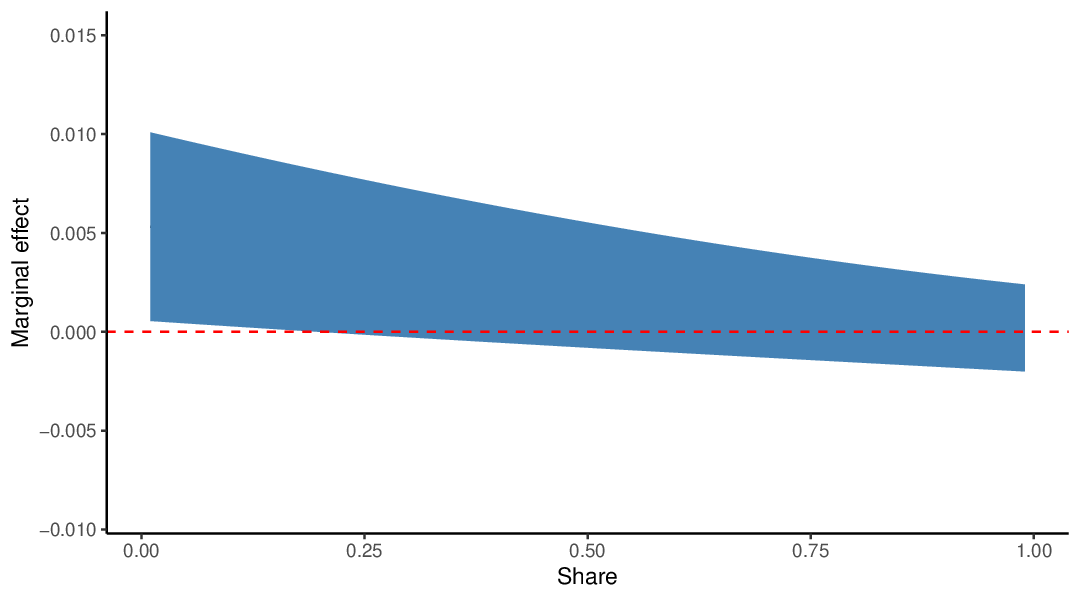}
		\caption{Emigration}
		\label{marginal_emigrated}
	\end{subfigure}
	\hfill
	\begin{subfigure}[b]{0.49\textwidth}
		\includegraphics[width=\textwidth]{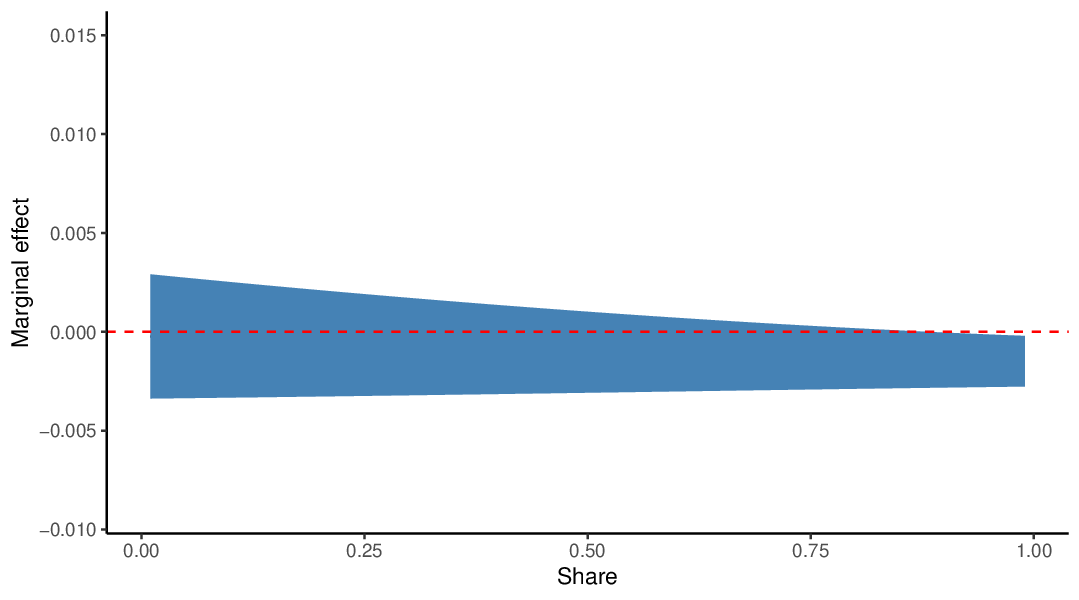}
		\caption{Share of foreign-born co-workers}
		\label{marginal_coworkers}
	\end{subfigure}
	
	\textit{Notes:} The plots display marginal effects based on estimates from a cubic specification that includes program and cohort fixed effects in Model \ref{model_nonlinear}. Marginal effects are computed as the derivative of the fitted outcome with respect to the share of international students, and standard errors are obtained via the delta method. Shaded areas represent 95\% confidence intervals.
\end{figure}

\pagebreak

\begin{figure}[!htb]
\caption{Marginal effects of the share of international students}
\label{marginal_2}
	\begin{subfigure}[b]{0.49\textwidth}
		\includegraphics[width=\textwidth]{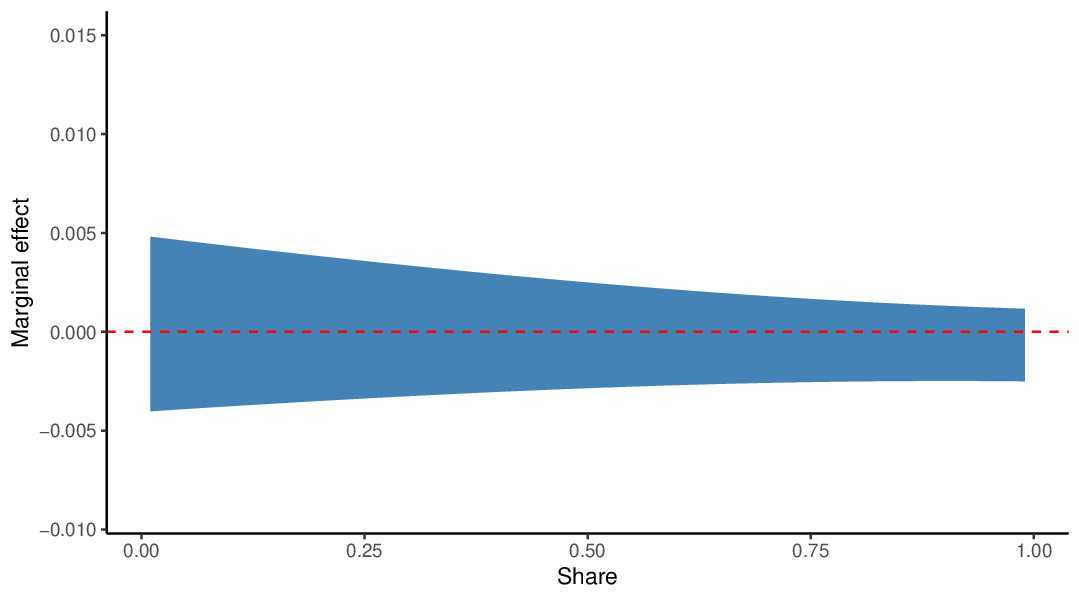}
		\caption{Employment}
		\label{marginal_employed}
	\end{subfigure}
	\hfill
	\begin{subfigure}[b]{0.49\textwidth}
		\includegraphics[width=\textwidth]{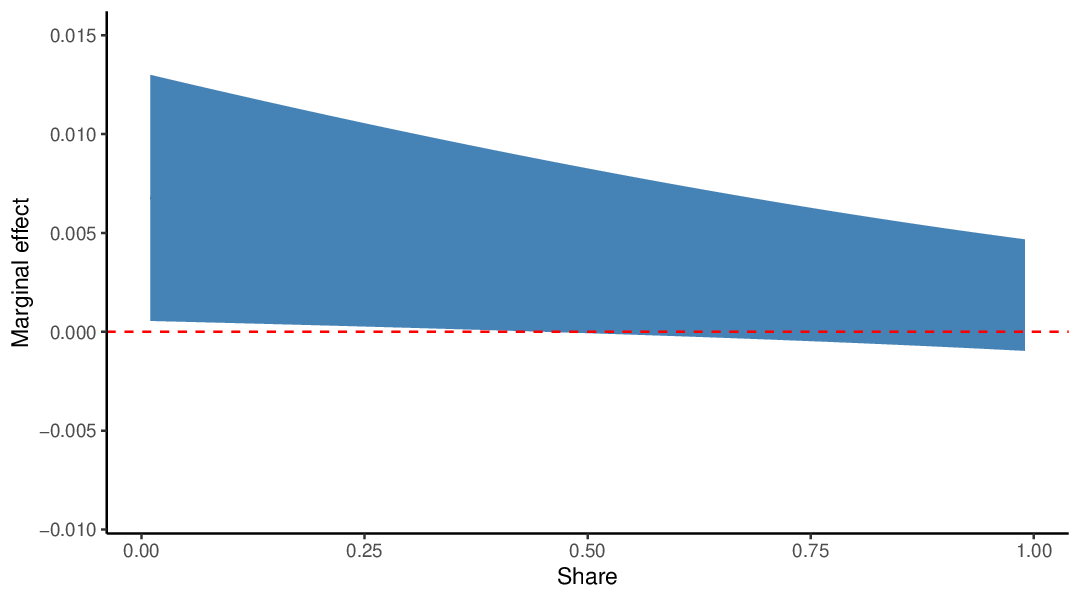}
		\caption{Income percentile}
		\label{marginal_income}
	\end{subfigure}
	
	\vspace{1em}
	
	\begin{subfigure}[b]{0.49\textwidth}
		\includegraphics[width=\textwidth]{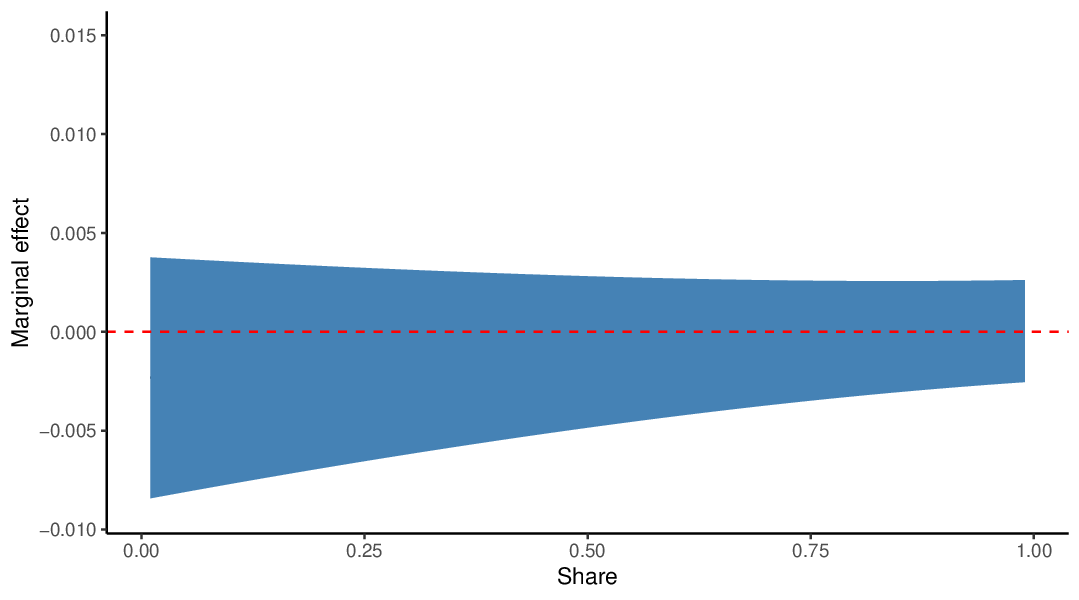}
		\caption{Entrepreneur}
		\label{marginal_entrepreneur}
	\end{subfigure}
	\hfill
	
	\textit{Notes:} The plots display marginal effects based on estimates from a cubic specification that includes program and cohort fixed effects in Model \ref{model_nonlinear}. Marginal effects are computed as the derivative of the fitted outcome with respect to the share of international students, and standard errors are obtained via the delta method. Shaded areas represent 95\% confidence intervals.
\end{figure}

\pagebreak

\begin{figure}[!htb]
\caption{Counterfactual treatment effects}
\label{placebo_1}
    \begin{subfigure}[b]{0.49\textwidth}
        \centering
        \includegraphics[width=\textwidth]{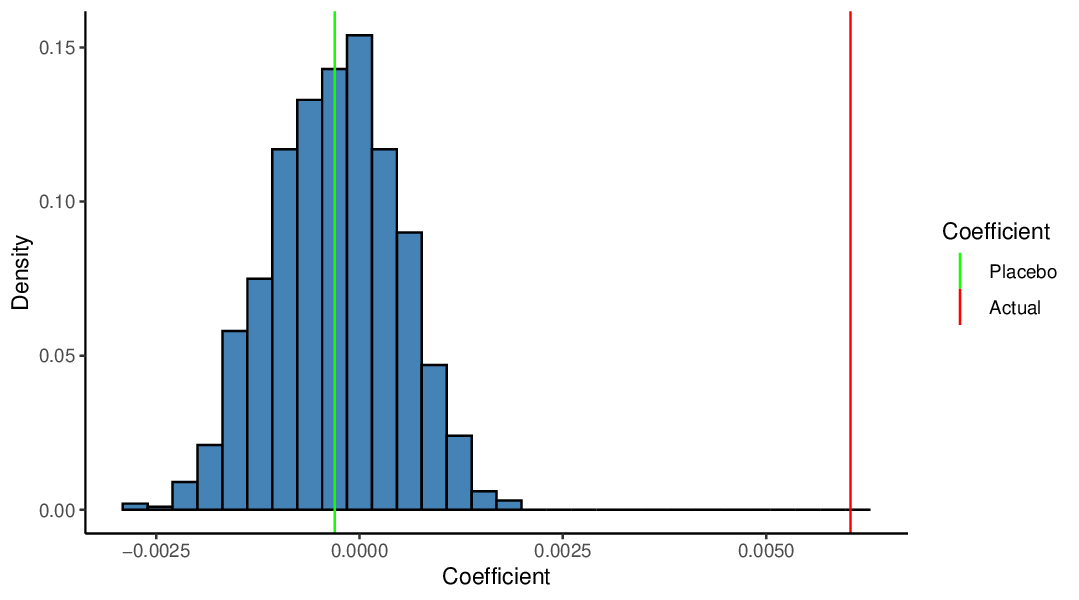}
        \caption{Cohabitation with a non-native}
        \label{placebo_cohabited}
    \end{subfigure}
    \hfill
    \begin{subfigure}[b]{0.49\textwidth}
        \centering
        \includegraphics[width=\textwidth]{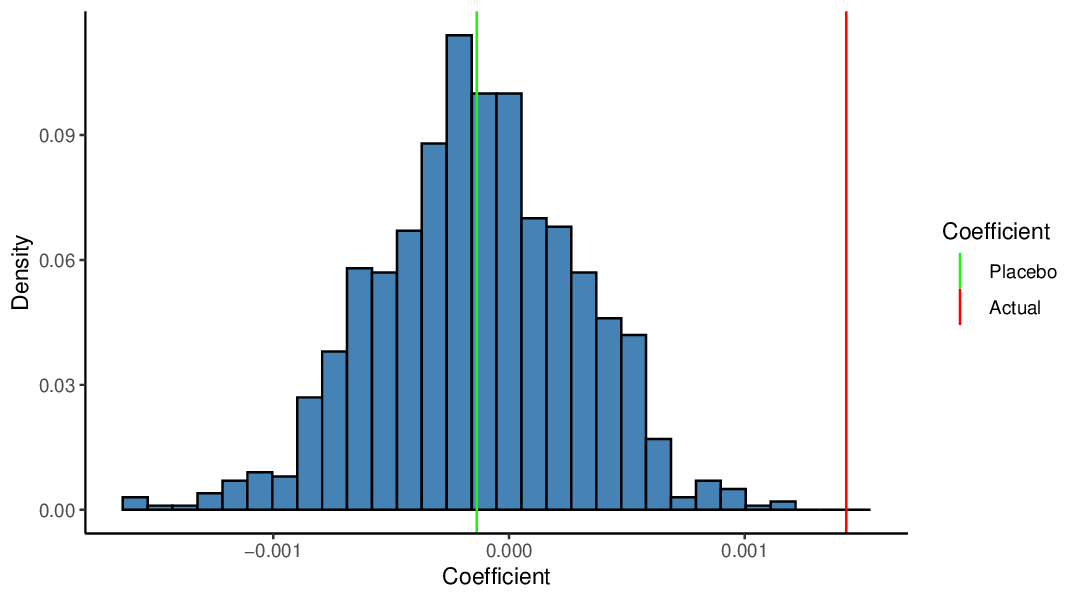}
        \caption{Marriage to a non-native}
        \label{placebo_married}
    \end{subfigure}
    
    \vspace{1em}
    
    \begin{subfigure}[b]{0.49\textwidth}
        \centering
        \includegraphics[width=\textwidth]{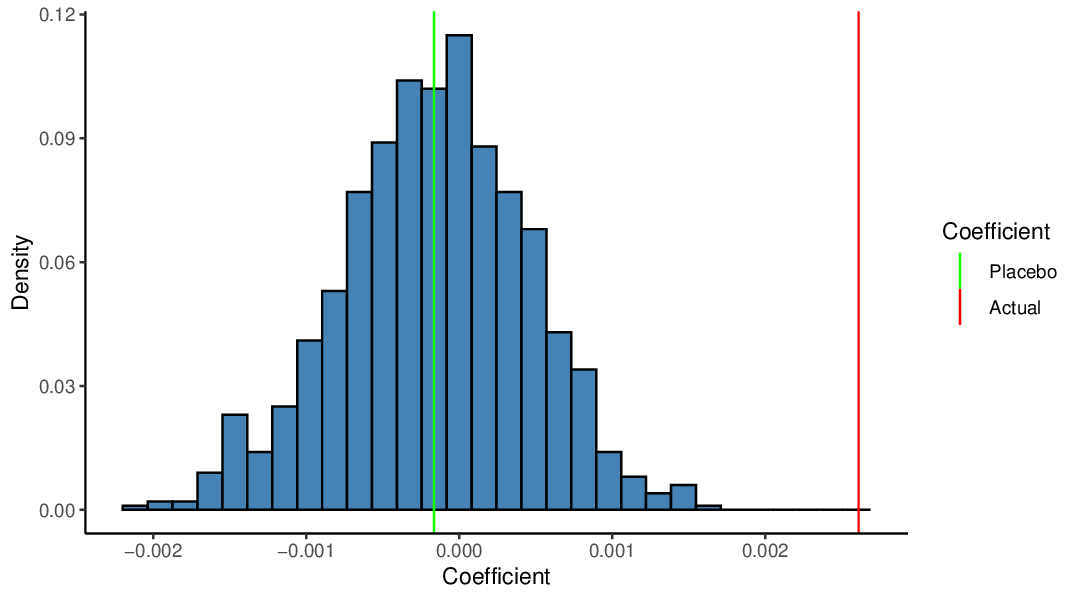}
        \caption{Emigration}
        \label{placebo_emigrated}
    \end{subfigure}
    \hfill
    \begin{subfigure}[b]{0.49\textwidth}
        \centering
        \includegraphics[width=\textwidth]{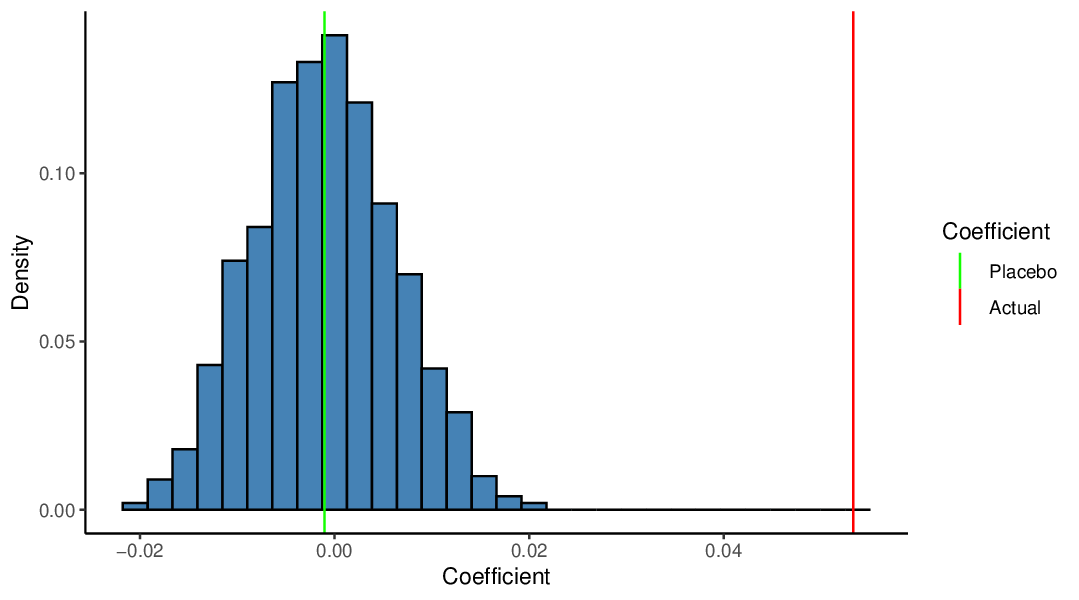}
        \caption{Satisfaction with internationalization}
        \label{placebo_international}
    \end{subfigure}
    
    \textit{Notes:} The plots display estimates from 1,000 simulated datasets where instead of using the actual share of international students that native students are exposed to, I assign them a share of international students from the same program and year but a different university. They also show the mean estimated coefficients of the randomly assigned shares (green), and the actual estimated coefficient (red). All regressions include program and cohort fixed effects.
\end{figure}
\pagebreak

\begin{figure}[!htb]
\caption{Counterfactual treatment effects}
\label{placebo_2}
    \begin{subfigure}[b]{0.49\textwidth}
        \centering
        \includegraphics[width=\textwidth]{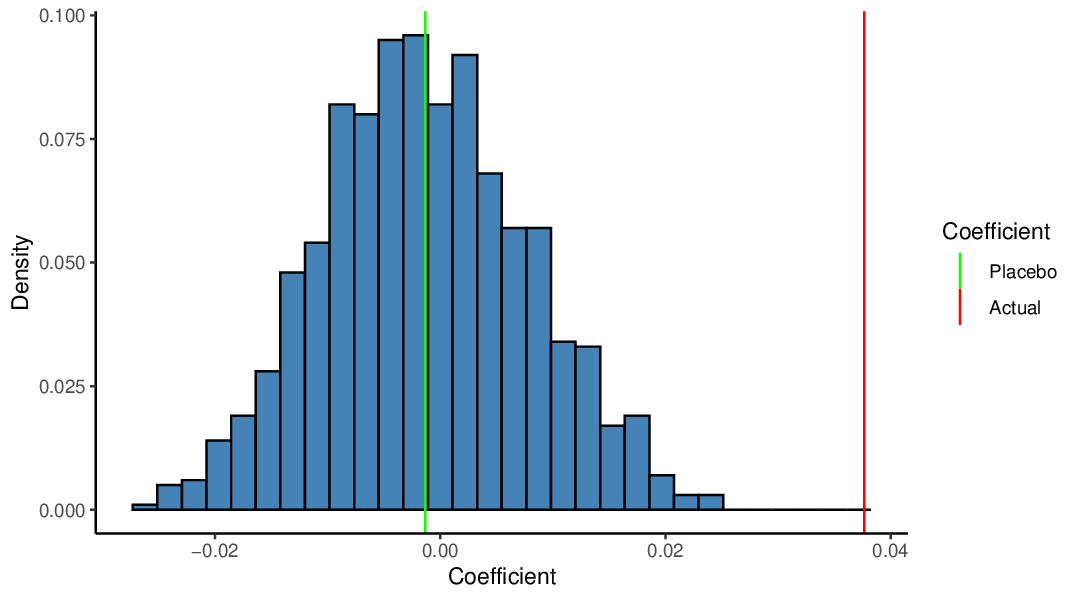}
        \caption{Satisfaction with encouragement to learn about other cultures}
        \label{placebo_international_2}
    \end{subfigure}
    \hfill
    \begin{subfigure}[b]{0.49\textwidth}
        \centering
        \includegraphics[width=\textwidth]{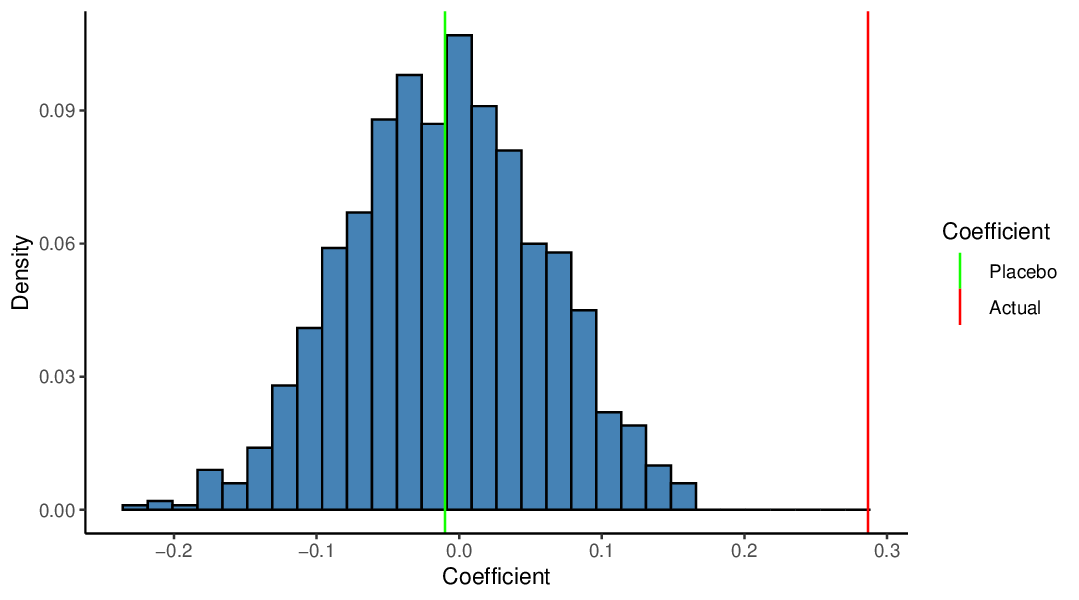}
        \caption{Same social security rights for foreigners}
        \label{placebo_social_rights}
    \end{subfigure}
    
    \vspace{1em}
    
    \begin{subfigure}[b]{0.49\textwidth}
        \centering
        \includegraphics[width=\textwidth]{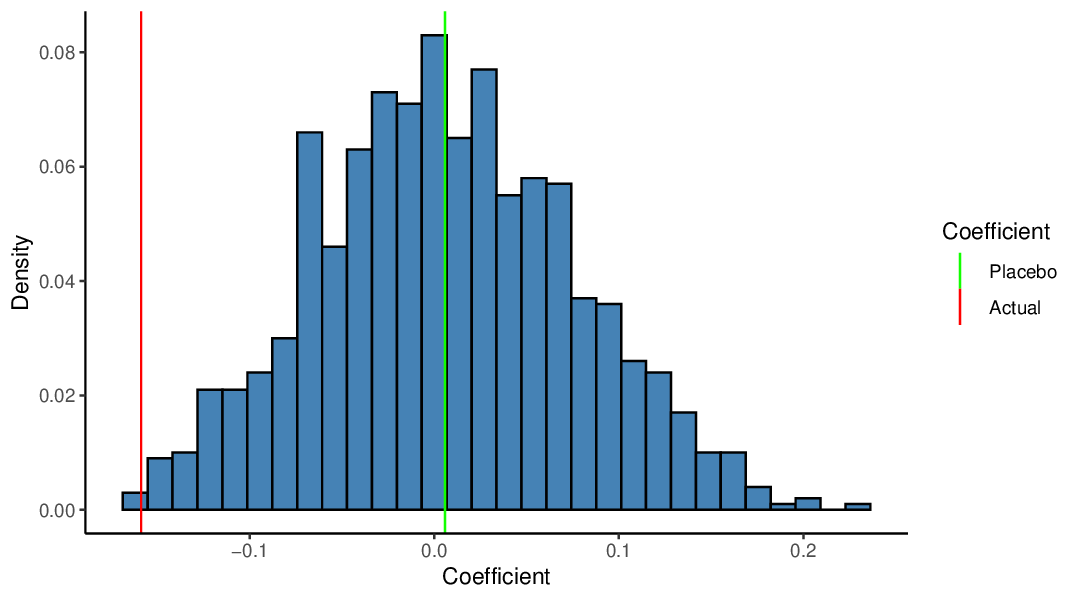}
        \caption{Foreigners in neighborhoods}
        \label{placebo_neighborhood}
    \end{subfigure}
    \hfill
    \begin{subfigure}[b]{0.49\textwidth}
        \centering
        \includegraphics[width=\textwidth]{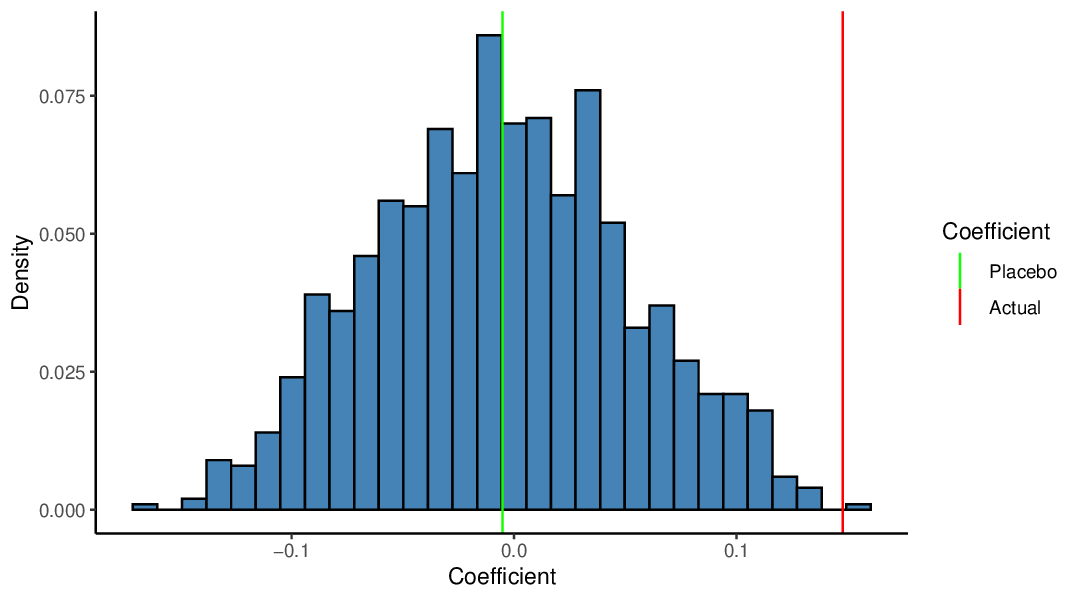}
        \caption{European unification}
        \label{placebo_european}
    \end{subfigure}
    
    \textit{Notes:}  The plots display estimates from 1,000 simulated datasets where instead of using the actual share of international students that native students are exposed to, I assign them a share of international students from the same program and year but a different university. They also show the mean estimated coefficients of the randomly assigned shares (green), and the actual estimated coefficient (red). All regressions include program and cohort fixed effects.
\end{figure}
\pagebreak

\begin{figure}[!htb]
\caption{Counterfactual treatment effects}
\label{placebo_3}
    \begin{subfigure}[b]{0.49\textwidth}
        \centering
        \includegraphics[width=\textwidth]{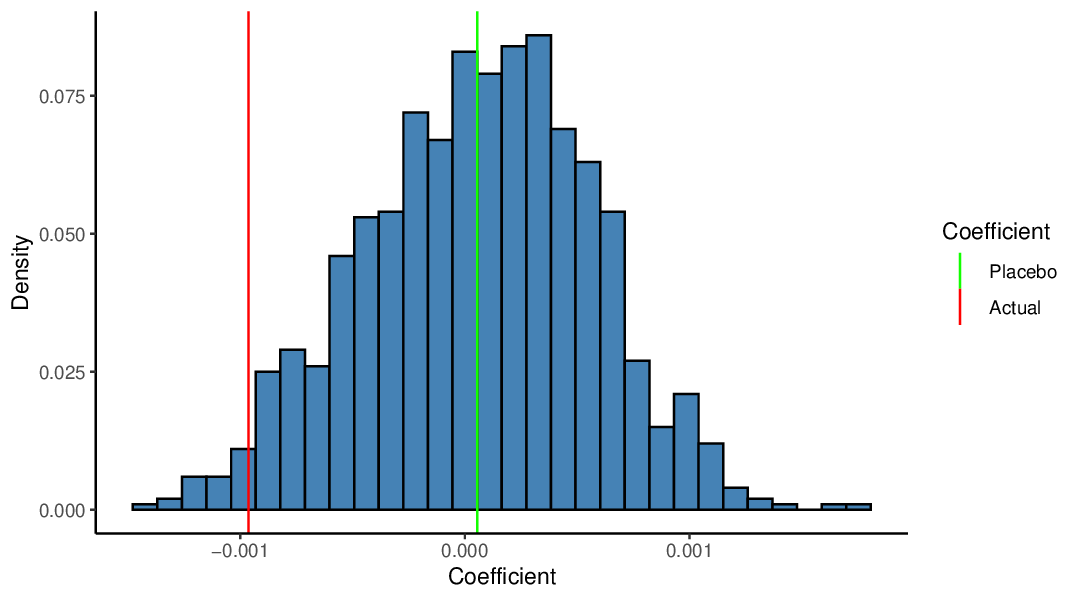}
        \caption{Employment}
        \label{placebo_employed}
    \end{subfigure}
    \hfill
    \begin{subfigure}[b]{0.49\textwidth}
        \centering
        \includegraphics[width=\textwidth]{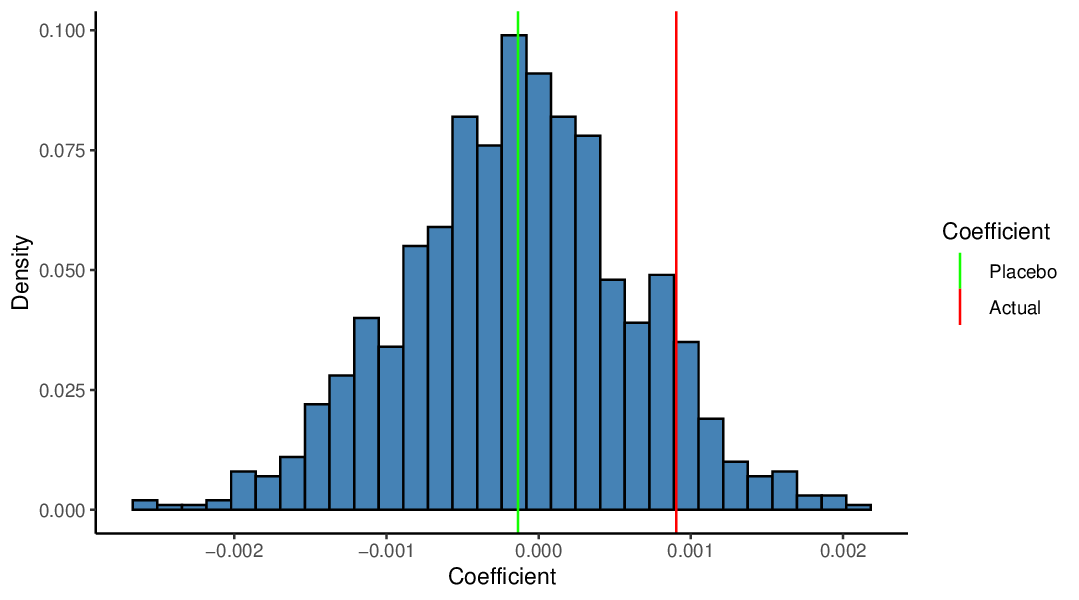}
        \caption{Income percentile}
        \label{placebo_income}
    \end{subfigure}
    
    \vspace{1em}
    
    \begin{subfigure}[b]{0.49\textwidth}
        \centering
        \includegraphics[width=\textwidth]{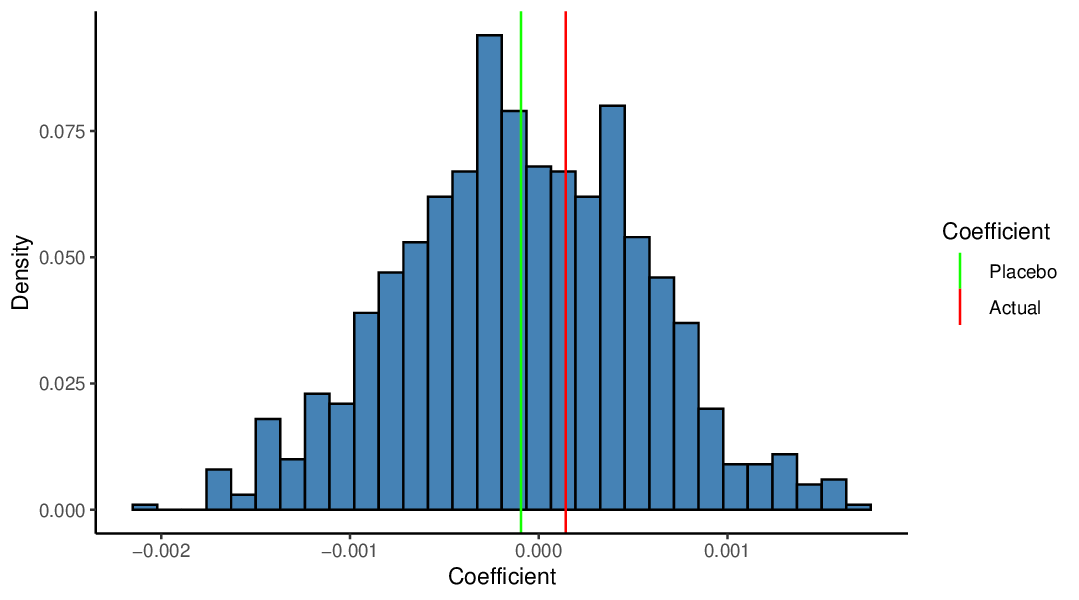}
        \caption{Entrepreneur}
        \label{placebo_entrepreneur}
    \end{subfigure}
    \hfill
    \begin{subfigure}[b]{0.49\textwidth}
        \centering
        \includegraphics[width=\textwidth]{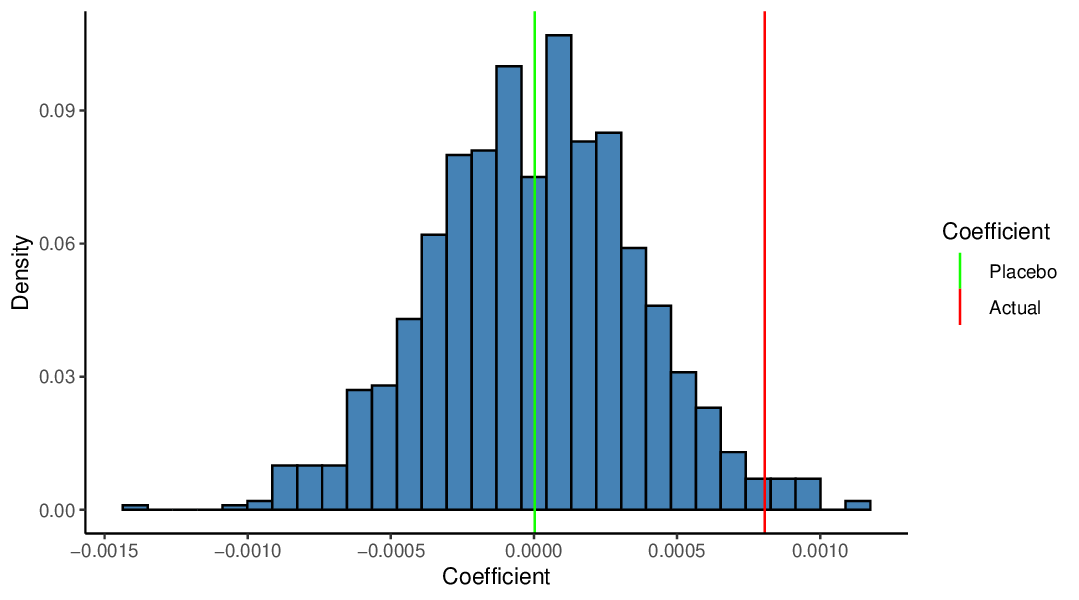}
        \caption{Share of foreign-born co-workers}
        \label{placebo_coworkers}
    \end{subfigure}
    
    \textit{Notes:}  The plots display estimates from 1,000 simulated datasets where instead of using the actual share of international students that native students are exposed to, I assign them a share of international students from the same program and year but a different university. They also show the mean estimated coefficients of the randomly assigned shares (green), and the actual estimated coefficient (red). All regressions include program and cohort fixed effects.
    \label{group3}
\end{figure}

\end{document}